\documentclass[preprint,authoryear,12pt]{elsarticle}

 \usepackage{graphicx}

\usepackage{amssymb}
\usepackage{amsmath}
\usepackage{algorithm}
\usepackage{algorithmic}

\journal{Journal of Comp.\,Statistics \& Data Analysis}

\begin{document}

\begin{frontmatter}



\title{Directional Variance Adjustment: improving covariance estimates for high-dimensional portfolio optimization
}



\author[tu,qpl]{Daniel Bartz\corref{cor1}}
\ead{daniel.bartz@tu-berlin.de}

\author[db]{Kerr Hatrick}

\author[db]{Christian W.~Hesse}

\author[tu]{Klaus-Robert M\"uller}

\author[tu]{Steven Lemm}


\address[tu]{Machine Learning Group, Computer Science Dept., TU Berlin, Berlin}

\address[db]{Global Markets Equity, Deutsche Bank AG, London}

\begin{abstract}
Robust and reliable covariance estimates play a decisive role in
financial and many other applications. An important class of estimators is based on
Factor models. Here, we show by extensive Monte Carlo simulations that covariance matrices derived from
the statistical Factor Analysis model exhibit a systematic error, which is similar to the
well-known systematic error of the spectrum of the sample covariance
matrix. Moreover, we introduce the \emph{Directional Variance
Adjustment (DVA)} algorithm, which diminishes the systematic error. In
a thorough empirical study for the US, European, and Hong Kong market
we show that our proposed method leads to improved portfolio
allocation.

\end{abstract}

\begin{keyword}
covariance estimation \sep factor models \sep portfolio optimization


\end{keyword}

\end{frontmatter}

\newcommand\transpose{^\top}
\newcommand{\upth}[1]{\ensuremath{ #1^{\mbox{\tiny th}}}}

\newcommand{\comment}[2][??]{\textbf{[#1: #2]}}
\newcommand{\SL}[1]{\comment[SL]{#1}}

\renewcommand{\algorithmicrequire}{\textbf{Input:}}
\renewcommand{\algorithmicensure}{\textbf{Output:}}

\section{Introduction and Motivation}
The advent of modern finance began with Markowitz and his seminal
paper on portfolio optimization (\cite{Mar52}).  His theory provides a
mathematical approach to diversification by directly minimizing the
portfolio variance. Moreover, by adding constraints to the
optimization problem, we can e.\,g.\,prohibit or allow
short-selling. Other applications comprises the creation of portfolios
which constitute optimal hedges or track indices.  However, a
fundamental issue in portfolio allocation is the accurate and precise
estimation of the covariance matrix of asset returns from
historical data.

Covariance estimation and coping with its uncertainties have occupied
both researchers and practitioners since then. One of the major
difficulties with robust covariance matrix estimation arises from
nonstationarity of financial time series (see,
e.g. \cite{LorPhi94}, \cite{PagSch90}). Here, changes in the data
generating processes force the estimation to rely on short time
windows of recent observations. On the other hand the number of
parameters increases quadratically with the number of assets, i.e.,
for a set of $N$ assets, the covariance matrix has $\frac{1}{2}N(N+1)$
free parameters. For example, in order to estimate the covariance
matrix from daily return series of a moderately sized universe of one
hundred assets, already 5050 free parameters have to be
estimated. Following a general rule of thumb, that 10 observations per
parameter are required for a reliable estimate, the observation window
would need to cover approximately two years of data. Such a temporal horizon,
however, clearly contradicts with reported nonstationarity of
financial time series.  In practice, the situation is even exacerbated
by non-Gaussianity of financial time series\footnote{Return
time series often exhibit leptokurtic distributions.} (see,
e.g., \cite{LorPhi94}, \cite{Lon05}, \cite{Cam08}), which increases
the difficulty of covariance estimation even further, especially in case of small
sample sizes.  A possible remedy for problems caused by non-Gaussianity are robust estimation techniques
(\cite{Hub81}).

As the terms \emph{high dimensional} and \emph{small sample size} are rather vague and interdependent, the difficulty of the task of covariance estimation is commonly characterized by the ratio of sample size to dimensionality, $T/N$, which governs the properties of the spectrum of the sample covariance matrix (\cite{MarPas67,EdeRao05}).
For situations where this ratio is close to one or even below, many estimators which give better results than the sample covariance matrix have been proposed. Here, an important class is formed by regularized estimators, in which the
effective degrees of freedom are reduced by shrinkage (see,
e.\,g., \cite{Ste56,Fri89,LedWol03,SchStr05}). Another way to reduce
the degrees of freedom is to impose a latent structure on the
data. Here, commonly factor models (FMs) are in use. FMs assume the data
to be generated as a mixture of a small number of factors with
additive noise (\cite{Fan08}, \cite{GolIye03}).

In this paper, we will analyse a purely statistical factor model called (Maximum Likelihood)
Factor Analysis (see, e.\,g., \cite{Bas94}). 
As there is no analytic solution for the parameters of the Factor Analysis model,
we cannot provide a stringent
theoretic analysis of its properties. Instead, by means of thorough
simulations, we will provide evidence that the spectrum of the
covariance matrix derived from a Factor Analysis model is
biased\footnote{Here, we follow the terminology in \cite{Fri89}, who
deals with the bias in the spectrum of the sample covariance
matrix. We do not distinguish between bias and systematic error.}.  To
reduce the bias, we will propose the \emph{Directional Variance Adjustment
(DVA)} algorithm, which estimates the magnitude of the
imposed bias in specific directions by means of a Monte Carlo sampling approach and hence
enables for its correction.

In the portfolio optimization literature Monte Carlo sampling is known from Resampling Efficiency (\cite{Mic98}). There, the authors follow a  fundamentally different approach. While we use resampling to reduce the bias of our factor model, in Resampling Efficiency the sample mean and covariance are used to generate additional data sets, on which optimal portfolio weights are calculated which are then averaged. This is supposed to lead to more stable and diversified portfolios, but there is an ongoing debate on the merits of this procedure (see, e.g. \cite{Sch04}). Though not based on Monte Carlo resampling, techniques for the correction of variance inflation in principal components analysis are more related to our algorithm (\cite{Kje01,Abr11}).

At this point we would like to emphasize that in this paper we will solely focus
on the structure of risk in the stock market. A discussion about the
structure of expected returns (see, e.\,g.~$\beta$-pricing
models, \cite{Sha92}) is not within the scope of the paper.

We will evaluate our novel covariance estimation
procedure in the context of portfolio optimization, where we will compare
the proposed DVA Factor Analysis model to the sample covariance, Resampling Efficiency, Shrinkage,
standard Factor Analysis and the Fama-French Three-Factor model
(\cite{FamFre92}). By means of analyzing daily return data from
2001--2009 of three different markets, namely the US, EU and Hong Kong
stock markets, we will show that our proposed covariance matrix
estimation scheme leads to an improved portfolio allocation and hence
provide evidence that it better reflects the market's risk structure.

The paper is organized as follows. Section \ref{sec:covest} reviews
covariance estimation methods. In section \ref{sec:methods}, we will review
Factor Analysis and 
 investigate the bias in Factor Analysis by means of simulated
data.  Then, we will introduce our novel DVA approach for dealing with
the systematic error in the model and show the effectiveness in additional simulations.
In Section \ref{sec:empirics} we will present the results of a thorough comparative
study of various covariance estimation methods in the context of portfolio
optimization. Section \ref{sec:discussion} concludes the paper.

\section{Covariance Estimation}
\label{sec:covest}
\subsection{Sample Covariance Matrix and Systematic Error in its Spectrum}
\label{sec:sc}
The sample covariance matrix, 
\begin{align}
C^{sc}_{ij} = \frac{1}{T - 1} \sum^T_{t=1} (r_{ti} - \bar{r_i}) \cdot (r_{tj} - \bar{r_j}), \label{eq:sc}
\end{align}
where $\mathbf{R}$ is the ($T \times N$)-matrix containing $T$
observations of $N$ variables, is a consistent estimator of the
covariance matrix. This means that for $T \rightarrow \infty$ the sample
covariance matrix converges to the true covariance matrix.  When the ratio T/N is not large, however, the sample covariance matrix tends to be
ill-conditioned, implying that its inverse incurs large errors. In the
extreme case, when the number of observations falls below the number
of variables, the covariance matrix gets singular.

\begin{figure} 
\begin{center}
\includegraphics[width= 0.8 \linewidth]{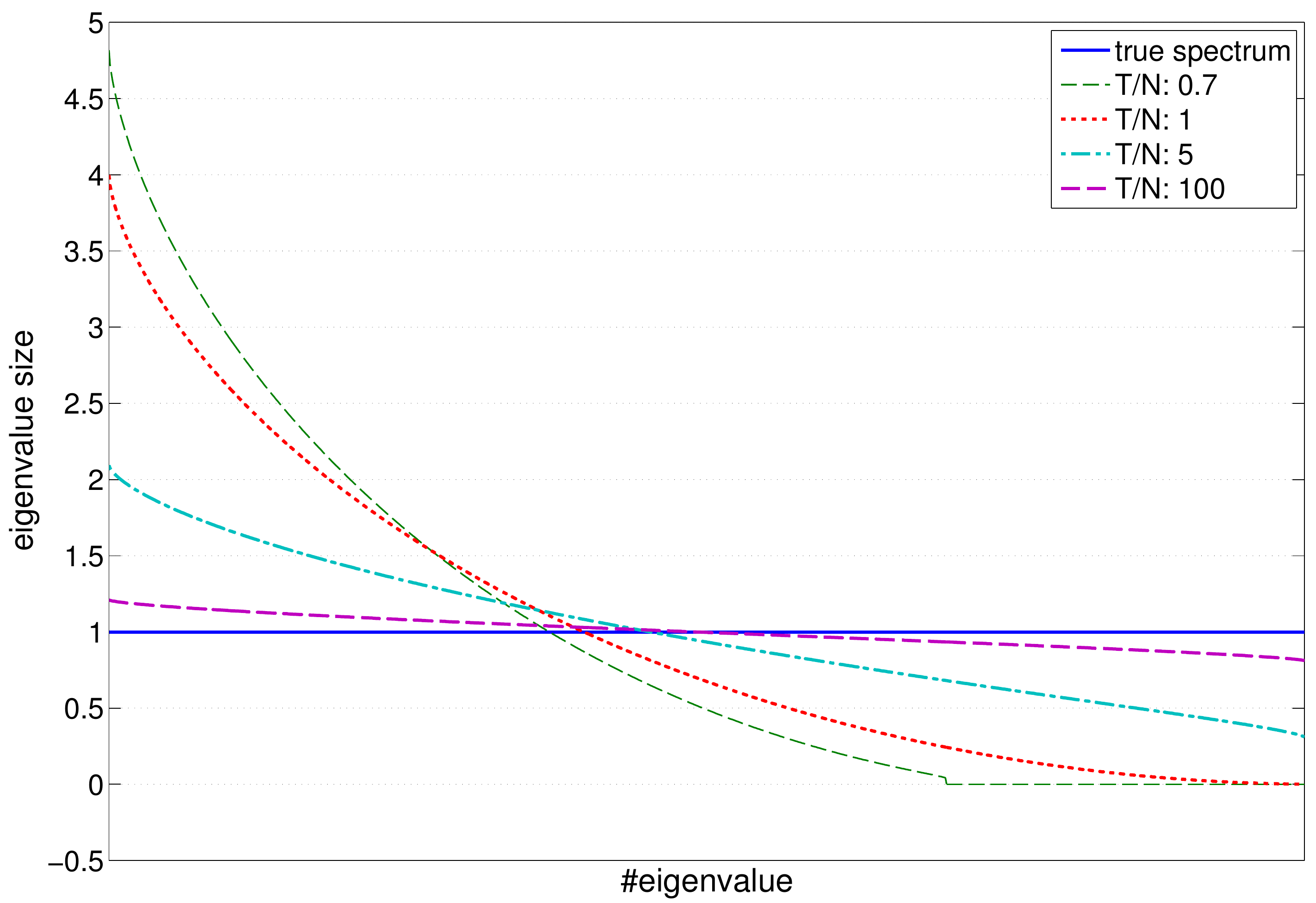}
\caption{Systematic error in the estimated eigenvalues of the sample
  covariance matrix for different ratios of sample size to dimensionality.} \label{fig:scbias}
\end{center}
\end{figure}

\label{sec:scbias}
Though the sample covariance is an unbiased estimator of the true
covariance matrix, this estimator exhibits a
systematic misestimation of the spectrum of the covariance
matrix which depends on the ratio of observations to dimensionality $T/N$. In particular, large and small Eigenvalues are
systematically over- and underestimated, respectively (see, e.\,g.~\cite{Fri89}).  
In order to illustrate this systematic error, we generated empirical spectra from the Mar\v{c}enko-Pastur density of eigenvalues for i.i.d.\,standard normally distributed variables (\cite{MarPas67}). The Mar\v{c}enko-Pastur density is the eigenvalue density in the limit $T,N \rightarrow \infty$, but already for sample sizes as small as 20 or 30 the empirical distribution is very similar (\cite{TulVer04}).
Figure~\ref{fig:scbias} shows the analytical solution for the empirical spectra  for various ratios of sample sizes to dimensionality. The magnitude of the
systematic error scales with the inverse of this ratio, for the
degenerate case ($T/N < 1$) there are $N-T$ zero eigenvalues. Even for
$T/N = 100$, the spectrum still differs visibly from the
true one.

Several methods have been proposed in the literature for correcting
the spectrum.  In Shrinkage (\cite{LedWol03,LedWol04,SchStr05}), the
goal is to find a suitable convex combination of the sample covariance
matrix $\mathbf{C}^{sc}$ and a shrinkage target $\mathbf{C}^{target}$,
\begin{align}
\mathbf{C}^{sh} = \lambda \mathbf{C}^{sc} + (1 - \lambda) \mathbf{C}^{target}, \label{eq:shrinkage}
\end{align}
where the shrinkage target is either fixed
(e.\,g.\,$\mathbf{C}^{target} = \mathbf{I}$) or a biased estimator
with lower variance (e.\,g.\,all correlations set to their average value).  For selecting the optimal shrinkage strength~$\lambda$, \cite{LedWol04}  proposed an analytic solution, which is computationally faster than the commonly used model selection via crossvalidation.
 Shrinkage can
be combined with factor modelling by taking a factor model as the
shrinkage target (\cite{LedWol03}).

Random Matrix Theory (RMT, for an overview see \cite{EdeRao05}) allows for several alternative
approaches to correct the spectrum. \cite{Ros02} propose to retain only those eigenvalues of the correlation matrix which are larger than the largest eigenvalues of a random matrix, given by the Mar\v{c}enko-Pastur law, and therefore likely to reflect some real structure. The model itself is equivalent to a PCA factor model based  on the correlation matrix, where RMT is used for selection of the appropriate number of factors. A similar model is proposed by \cite{Lal00}. Instead of setting the eigenvalues in the bulk of the spectrum to zero, they are set to their average value. A detailed analysis of these methods is beyond the scope of this article. Note that these methods are closely related to the PCA and the FA factor model which we will discuss. Thus, these models exhibit a similiar performance and suffer from the same bias we discuss in the following.

An interesting approach is described in \cite{Kar08}. There, the Mar\v{c}enko-Pastur law which describes the distribution of the sample eigenvalues is inverted numerically in order to obtain the true spectrum from the sample. For this, one has to be aware of two facts: first, the inversion is not unique and therefore a  prior or parametric ansatz has to be applied. Second, the largest eigenvalue of the covariance matrix of asset returns  is normally isolated from the bulk. This is problematic, because the inversion leads to a continous spectrum. These aspects make the application of this approach less straightforward and, to our knowledge, no publication with portfolio simulations exists in which a competetive performance was achieved.

The following section introduces Factor Models as a type of restricted
covariance estimator.

\subsection{Factor Models as Restricted Covariance Estimators}
\label{sec:fm}

In finance, factor models form an important class of restricted
covariance estimators. In a factor model, the returns $r_{ti}$ of the
\upth{i} asset at time $t$ are described as a weighted sum of $M$
random factor returns $f_{tm}$ multiplied with exposures $X_{mi}$ to these
factors and additional random noise $e_{ti}$:
\begin{align}
r_{ti} =  \; &\underbrace{\sum_{m=1}^{M}  f_{tm} \cdot X_{mi}}_{systematic \; risk} \quad + \underbrace{\; \; \; e_{ti}  \; \;}_{specific \; risk} \label{eq:fmc} \\
& e_i{\perp\!\!\!\perp} f_j,  \qquad \forall \; i,j       \notag \\
&  e_i{\perp\!\!\!\perp} e_j, \qquad \forall \; i \neq j  \notag .
\end{align}
Here, the systematic risk entirely describes the dependencies between
the assets, while the asset specific risks are assumed to be independent.
\begin{figure}[ht]
\begin{center}
\includegraphics[width= 0.6 \linewidth]{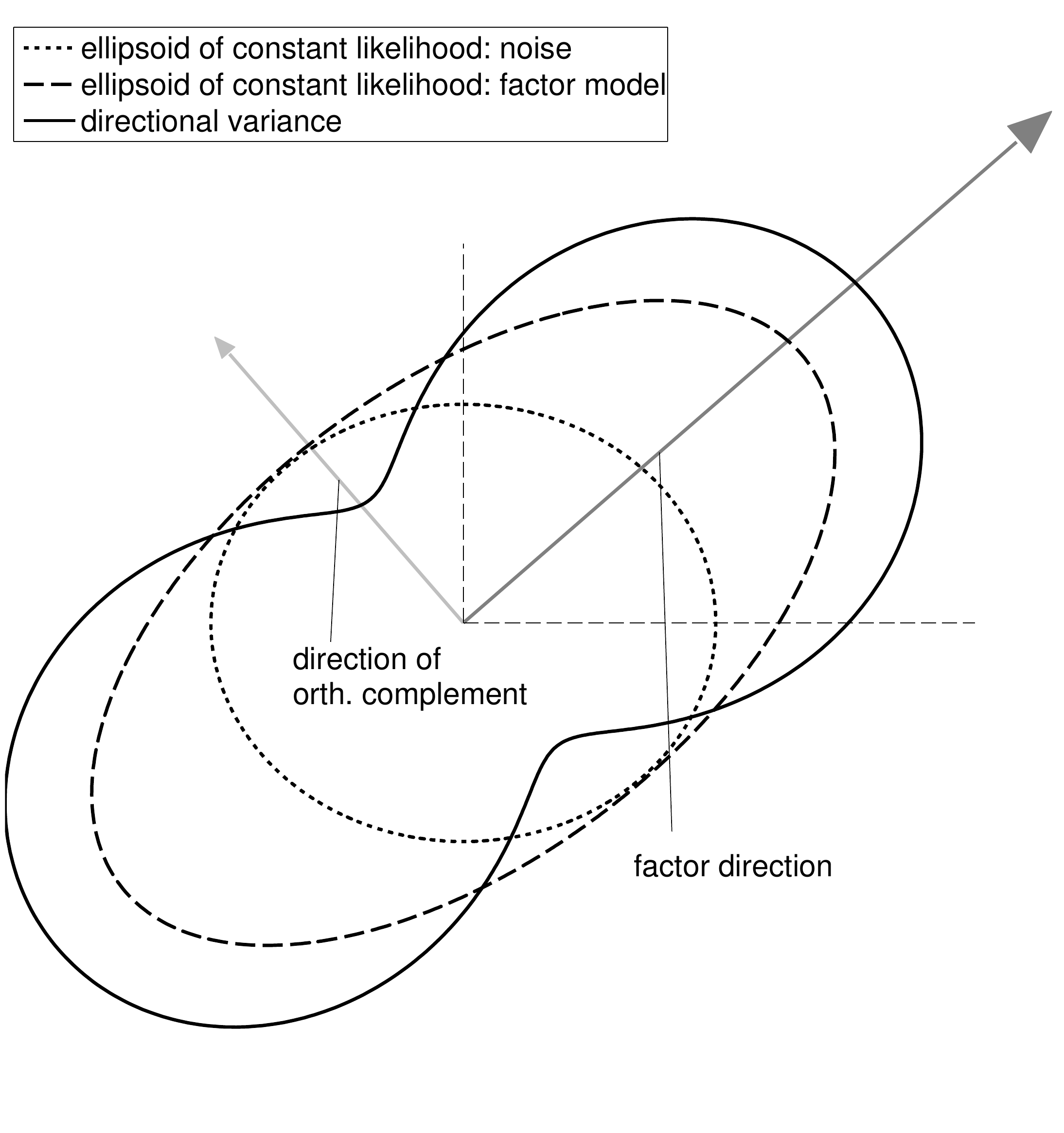}
\caption{A two dimensional example of a $1$-factor model. The arrows show the direction of the single factor and the orthogonal complement. The
  covariance matrices of the factor model $\mathbf{C}^{fm}$ (dashed) and the
  uncorrelated noise $\mathbf{\Sigma}_e $ (dotted) are shown as ellipsoids of constant
  likelihood. The peanut-shaped solid line shows the directional
  variances (${\bf v}\transpose\mathbf{C}^{fm} {\bf v}$) of the factor
  model along all directions ${\bf v}: \lVert{\bf v}\rVert_2=1$.} \label{fig:facovmat}
\end{center}
\end{figure}

In the statistics and signal processing literature, this is often
referred to as a mixture model, where $\mathbf{X}$ is the mixture
matrix and $\mathbf{f}$ are the source signals (see, e.\,g.,\cite{HyvOja00}. Calculating the
covariance matrix, one obtains
\begin{align}
\mathbf{C}^{fm} = \mathbf{R} \transpose \mathbf{R} = & \mathbf{(F X)} \transpose \mathbf{(F X)} + \mathbf{E}\transpose \mathbf{E} \notag \\
= & \mathbf{X}  \transpose \mathbf{\Sigma}_f \mathbf{X} + \mathbf{\Sigma}_e \label{eq:fmcov},
\end{align}
where $\mathbf{\Sigma}_f$ is the covariance of the factors and the
diagonal matrix $\mathbf{\Sigma}_e$ is formed by the asset specific
noise variances (cf.~Figure~\ref{fig:facovmat}). 

The advantage of factor models lies in the reduced number of
parameters for covariance estimation. Essentially, this means that a
higher bias is accepted in exchange for a reduced variance. In
quantitative finance, three different types of factors are employed to build up factor models:
fundamental, macroeconomic and statistical factors (\cite{Con95,Con10}).

In a fundamental factor model, assets are analysed and certain key metrics are used for setting up the factor model. 
Fundamental factor models are especially well suited when only a short history of  data is available, e.\,g.\;for weekly or monthly data, as fewer parameters have to be estimated from the history than in a statistical factor model.
The best-known model of this kind is the Fama-French
three-factor model (\cite{FamFre92}), in which the factor time series 
$\mathbf{f}$ are based on portfolios governed by market beta, book-to-market ratio and
market capitalization. The exposures to these factors are obtained from the
coefficients of a linear regression model.

In contrast, macroeconomic factor models predetermine the factors as
macroeconomic time series which are supposed to affect the asset
returns. As in the Fama-French model, the exposures  are obtained by linear regression. Examples for macroeconomic
time series used in factor models are unemployment rate, GNP, FX or
interest rates. However, for daily or higher frequency stock market returns, macroeconomic
factor models are of limited use and therefore neglected in the following (for an overview, see \cite{Con10}). 

The third approach, statistical factor modelling, is purely data
driven and extracts the factors as well as the exposures from
historical asset time series. Representatives of statistical factor
models are Principal Component Analysis (PCA, \cite{Jol86}),
Probabilistic Principal Component Analysis (PPCA, \cite{TipBis99}),
Independent Component Analysis (ICA, \cite{Com94,HyvOja00}) as well as
Factor Analysis (FA, see section \ref{sec:fa}).

Hybridization allows for models which combine statistical, fundamental and/or macroeconomic factors (\cite{Con95,Mil06,Con10}). As long as the hybrid models contain statistical factors, our approach could be adapted to improve covariance estimation. 

\section{Directional Variance Adjustment of Factor Analysis}
\label{sec:methods}

\subsection{(Maximum Likelihood) Factor Analysis} 
\label{sec:fa}

Factor Analysis is a latent variable model which has its roots in
psychology and answers the question for the "best" explanation of the
observed data for a given number of factors (latent variables). Here,
"best" model refers to the model that maximizes the data likelihood. The
application of Factor Analysis to financial data was first introduced
 in order to test the Arbitrage Pricing Theory (\cite{RolRos80}).

Factor Analysis models the asset returns as a mixture of unobserved source signals with additive
noise. The signals and the noise are assumed to be i.i.d., zero-mean normally
distributed. Independence of the noise ($\rightarrow$~diagonal noise
covariance matrix) and independence of noise and factors
($\rightarrow$~covariance is a sum of factor and noise contributions)
are assumed (cf.\ eq.~(\ref{eq:fmc})). In addition, it is assumed that
scaling and correlation of the systematic risk are contained in the
mixing matrix ($\rightarrow$~standard normally distributed independent
factors). Hence, the model reads as
\begin{align}
\mathbf{r}_t =  & \mathbf{f}_t \cdot \mathbf{X} + \boldsymbol{\epsilon}_t , \label{eq:fa} \\
\mathrm{with} \qquad \mathbf{f}_t  \sim {\cal N}( 0 , \mathbf{I} ),  & \quad  \quad \boldsymbol{\epsilon}_t  \sim {\cal N}(0 , \mathbf{D}),\notag 
\end{align} 
where $\mathbf{D}$ is a diagonal matrix. The corresponding log-likelihood is obtained as
\begin{eqnarray}
L(\mathbf{X},\, \mathbf{D})
= 
\ln p( \mathbf{R},\, \mathbf{F} | \mathbf{X},\, \mathbf{D} ) 
&=& \sum_{t=1}^T \left\{ \ln p(\mathbf{r}_t | \mathbf{f}_t ,
\mathbf{X},\, \mathbf{D}) + \ln p( \mathbf{f}_t) \right\}
.
\end{eqnarray}
Especially in the finance context, normality is a strong assumption. In order to make the model more appropriate for financial data, it is possible to extend FA to t-distributions (t-FA, see \cite{McL07}).  t-FA has the same bias as standard FA and our method can be adapted in a straightforward way by replacing FA by t-FA, but a comparison of these methods is beyond the scope of this paper.

We obtain estimates of the model parameters by
Expectation-Maximization\footnote{Different methods for solving the
  optimization problem are proposed in the literature. A popular
  alternative is based on the quasi-newton method (see
  \cite{Jor67}). As the algorithm described by J\"oreskog uses an
  eigendecomposition, which is costly to obtain in high dimensions
  (${\cal O}(N^3)$), we have opted for the EM approach (${\cal O}(M
  N^2)$). Other methods claiming superior performance suffer from the
  same drawback (see, e.g., \cite{Zha08}). Moreover, for the main
  claim of this paper, the  optimization procedure chosen
  to obtain the maximimum likelihood solution is of no importance.}
(EM, see \cite{Dem77}, for applications on Factor Analysis see
\cite{RubThay82} and \cite{Roweis99}).  In this algorithm, the
likelihood is maximized iteratively by alternating between the
Expectation and the Maximization step:
\begin{itemize}
\item in the Expectation step, the exposure $\mathbf{X}$ and noise
  variance $\mathbf{D}$ are assumed to be fixed and the expected
  factor $\mathbf{F}$ (latent variables) can be derived directly.
\item in the Maximization step, the expected factors $\mathbf{F}$ are
  assumed to be fixed and the likelihood is maximized with respect to
  exposures $\mathbf{X}$ and noise variances $\mathbf{D}$.
\end{itemize}
These two steps are iterated until convergence. The resulting
covariance matrix estimate of the Factor Analysis model is then given
as
\begin{align}
\hat{\mathbf{C}}^{fa} = \mathbf{\hat{X}} \transpose \mathbf{\hat{X}} +
\mathbf{\hat{D}}. \label{eq:FAcov}
\end{align}
Note that the above equation follows trivially from eq.~(\ref{eq:fmcov}) for independent and standard normal factors. For Factor Analysis the number of parameters is reduced from $\frac{1}{2}N(N+1)$ to
\begin{align}
df = &\; \underbrace{ M \cdot N}_{entries \, in \, X}  \quad - \underbrace{\; ( M - 1 )\;}_{rotational \, invariance\, of\, X} \quad + \underbrace{\; \;  N  \;}_{diagonal \, elements\, of\, D}  \notag \\
= & \; (M+1) \cdot (N-1) + 2.\label{eq:df}
\end{align}

\subsection{Systematic Error in Factor Analysis}
\label{sec:fabias}

Since there are no analytical results for the spectrum of Factor Analysis as there are for the sample covariance matrix (section \ref{sec:scbias}), we run a simulation to study systematic errors in Factor Analysis. To this end, we generate $N=30$ dimensional
return data according to an underlying three factor model as in
eq.~(\ref{eq:fa}). The noise covariance matrix $\mathbf{D}$ was
defined with equally spaced values from the intervall $[0.5, 1.5]$ on
the diagonal. The three rows of the mixing matrix $\mathbf{X}$ were generated as randomly oriented vectors with a length of 10, 3 and 1, respectively. 
In order to study the small sample size properties of
Factor Analysis for this setting, we set the ratio $T/N$ to 0.7, 1 and 5, corresponding to 21, 30, and 150 thirty-dimensional
observations. As $\mathbf{X}$ and $\mathbf{D}$ are known for the
simulation, the true covariance matrix $ \mathbf{C}^{true}$ can be
calculated by the population counterpart of  eq.\,(\ref{eq:FAcov}).


In section \ref{sec:scbias} we studied the systematic error of the
eigenspectrum of the sample covariance matrix, where the variance in the $i$-th eigendirection $\mathbf{v}_i$ corresponds to the size of the $i$-th eigenvalue $\lambda_i$:
\begin{align*}
\mathbf{v}_i\transpose \mathbf{C} \mathbf{v}_i = \mathbf{v}_i \transpose \lambda_i \mathbf{v}_i = \lambda_i.
\end{align*}
In the following we will study systematic errors in terms of
misspecification of directional variances. More precisely, we will
investigate systematic errors in the factor subspace and its
complementary orthogonal space separately. To this end we first
calculate an orthonormal basis $\mathbf{P}^0_{fs}$ ($N \times M$) of
the $M$-dimensional subspace in which the estimated factors
${\mathbf{\hat{X}}}$ lie (the \emph{Factor Subspace}) and another
orthonormal basis $\mathbf{P}^0_{oc}$ ($N \times (N-M)$) of the
$(N-M)$-dimensional orthogonal complement. Correspondingly, we can
confine the covariance matrix to the two subspaces, yielding a factor
space related part and its orthogonal counterpart as
\begin{align*} 
\mathbf{C}^{fa}_{fs} := & \; \mathbf{P}^0_{fs} {\mathbf{P}^0_{fs}}\transpose \mathbf{C}^{fa} \mathbf{P}^0_{fs} {\mathbf{P}^0_{fs}} \transpose \qquad \qquad \mathrm{and} \\
\mathbf{C}^{fa}_{oc} := & \; \mathbf{P}^0_{oc} {\mathbf{P}^0_{oc}}\transpose \mathbf{C}^{fa} \mathbf{P}^0_{oc} {\mathbf{P}^0_{oc}} \transpose.
\end{align*}
For each subspace, we obtain a new basis ($\mathbf{P}_{fs}$ and
$\mathbf{P}_{oc}$) as the corresponding  eigenbasis of
$\mathbf{C}^{fa}_{fs}$ and $\mathbf{C}^{fa}_{oc}$, respectively.
Combining these subspace bases\footnote{Here, we consider only the non-zero
  Eigenvalues and assume the Eigenvectors to be sorted in decreasing
  order with respect to their Eigenvalues.} to $\mathbf{P}$ = [$\mathbf{P}_{fs}$,$\mathbf{P}_{oc}$]
yields an orthonormal basis of the entire space ($\mathbb{R}^N$).

Along these directions $\mathbf{p}_i$ we measure the directional
variances $\sigma_i^2$ for the true and the estimated Factor Analysis
model and calculate the systematic error as 
\begin{align} 
S^{fa}_i = \mathbb{E} \left[ \frac{\sigma^{2  \, fa}_i}{\sigma^{2  \, true}_i}
  \right], \qquad \sigma^{2  \, true}_i = \mathbf{p}_i \transpose
\mathbf{C}^{true} \mathbf{p}_i \ ,  
\qquad
\sigma^{2  \, fa}_i = \mathbf{p}_i \transpose \mathbf{C}^{fa} \mathbf{p}_i. \label{eq:syserr}
\end{align} 
Here, values $S_i>1$ and $S_i<1$ correspond to an over- and
underestimation of the directional variances, respectively. Moreover,
the basis $\mathbf{P}$ explicitly takes the factor structure into
account. Hence, this particularly chosen basis enables us to study the
specific systematic estimation errors in the factor subspace and noise
subspace separately\footnote{The use of the conventional eigenbasis
of $\mathbf{C}^{fa}$ does not allow to disentangle the subspace
specific errors.}.  Note that the directions $\mathbf{p}_i$ are solely
derived from the estimated parameters of the factor model and do not rely on information
about the true covariance matrix.

\begin{figure} 
\begin{center}
\includegraphics[width=0.85\linewidth]{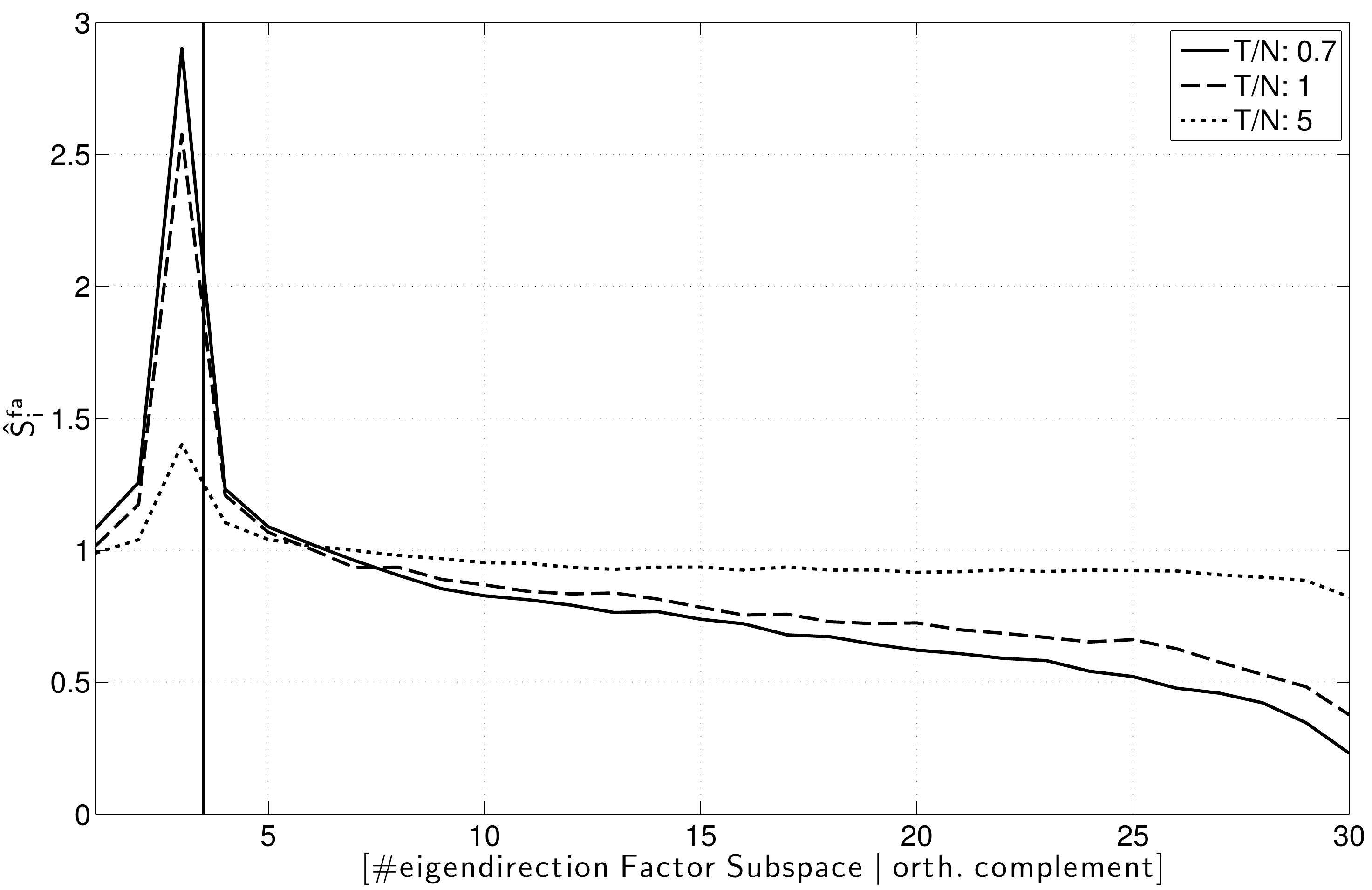}
\caption{average ratio between Factor Analysis and true variances in
  the factor subspace and the orthogonal complement. Ratios of sample
size to dimensionality $T/N=0.7$, $1$ and $5$. $N = 30$.  Average over 150 datasets.} \label{fig:fmbias2}
\end{center}
\end{figure}

Figure~\ref{fig:fmbias2} depicts the estimated systematic error $S$
(eq.~(\ref{eq:syserr})) of Factor Analysis by means of the simulated
data. Clearly, Factor Analysis tends to overestimate the variance in
the $3$-dimensional Factor Subspace, while the variance in the
orthogonal complement is on average underestimated. This is not
surprising, as the Factor Analysis model attributes strong covariances
in the sample to the factors. Consequently, factors with low
Signal-to-Noise-ratio (SNR) are hard to identify and directions of
spurious covariance are likely to be misrepresented as factors,
yielding an overestimating of the variance along these directions: In
the simulations, the strongest (first) factor, which has a high
Signal-to-Noise-Ratio can be estimated with very high accuracy even
for small sample sizes and the variance estimate does not have a
significant systematic error. The weaker factors with a lower SNR in
contrast tend to yield overestimated variances along the estimated
factor directions. This effect is highly pronounced for small sample
sizes and persists for relatively large sample sizes.

On the other hand, the noise subspace spectrum shows a similar --
albeit weaker -- behaviour as the spectrum of the sample covariance
matrix, i.e., variances corresponding to large eigenvalues are
overestimated, while variances corresponding to small
eigenvalues are underestimated (compare Figure~\ref{fig:scbias} and
Figure~\ref{fig:fmbias2}). As for the sample covariance matrix, this
effect is especially pronounced for small sample sizes.

\subsection{Directional Variance Adjustment: Correcting the Systematic Error}
The systematic error of the spectrum of a sample covariance matrix
with respect to the true spectrum can be estimated analytically: from
the distribution of the entries in the covariance matrix, the
distribution of the eigenvalues can be derived (see
e.g.,~\cite{EdeRao05}). The minimization of the Factor Analysis cost function on the other hand does not have a closed form solution,  an
iterative method has to be used. Hence it
does not facilitate an analytical approach to obtain the distribution
of the eigenvalues. Consequently, we will deploy a method that is
based on Monte-Carlo-sampling. 

To this end, suppose we have estimated the parameters $\cal{F}$ of a Factor Analysis model
 and want to correct the corresponding covariance
matrix $\mathbf{C}^{\cal{F}}$ for the systematic error. Then we
estimate the systematic error in the following manner: Using
${\cal{F}}$ for a generative model, we generate $K$ synthetic data sets
of the same size as the original sample. For each data set we estimate
a corresponding Factor Analysis parameter set ${\cal{F}}_1,\dots,{\cal{F}}_K$.
Note that for these parameter sets the true set of parameters (i.e., $\cal{F}$) is known
and with it the true covariance matrix. This enables us to quantify
the amount by which the directional variances along the
eigendirections of $\mathbf{C}_{fs}^{{\cal F}_k}$ (factor subspace)
and $\mathbf{C}_{oc}^{{\cal F}_k}$ (orthogonal complement) are over-
and underestimated, respectively. The estimated systematic errors, can
then directly be turned into multiplicative correction factors for the
adjustment of the directional variances of $\cal{F}$. Applying these
corrections to the eigendirections of the factor space and its
orthogonal complement yields to what we refer as
the \emph{directional variance adjusted covariance matrix}
$\mathbf{C}^{DVA}$ of $\cal{F}$ (see algorithm~\ref{alg:DVA}).
\begin{algorithm}[h!]
\caption{DVA}\label{alg:DVA}
\begin{algorithmic}[1]
\REQUIRE 
   the estimated parameters of the Factor Analysis model $\cal{F}$ ;
   the sample size $T$;
   the number of Monte Carlo runs  $K$   
\ENSURE the directional variance adjusted covariance $\mathbf{C}^{DVA}$
\medskip
\STATE generate $K$ synthetic data sets of size $T$ based on $\cal{F}$.
\STATE from the $K$ data sets, estimate $K$ factor model parameter sets ${\cal{F}}_1,\dots,{\cal{F}}_K$
\STATE For each ${\cal{F}}_k$, estimate the basis $\mathbf{P}_k$ = [$\mathbf{P}_{k,fs}$,$\mathbf{P}_{k,oc}$] (see sec. \ref{sec:fabias})
\STATE estimate the directional variance correction factors\\
\vspace{0.1cm}$\vspace{0.1cm} \hspace{3cm} S^C_i = \; \frac{1}{K} \sum_{k=1}^K  \frac{ \mathbf{p}_{k,i}\transpose \mathbf{C}^{{\cal{F}}_k} \mathbf{p}_{k,i}}{ \mathbf{p}_{k,i}\transpose \mathbf{C}^{\cal{F}} \mathbf{p}_{k,i}}$
\STATE For ${\cal{F}}$, estimate the basis $\mathbf{P}$ = [$\mathbf{P}_{fs}$,$\mathbf{P}_{oc}$]
\STATE calculate the directional variance adjusted covariance matrix\\ 
\vspace{0.1cm} $\vspace{0.1cm} \hspace{3cm} \mathbf{C}^{DVA} = \mathbf{C}^{\cal{F}} + \sum_{i=1}^N ( 1 - 1/S^C_i ) \cdot (\mathbf{p}_{i} \mathbf{p}_{i}\transpose)  \mathbf{C}^{\cal{F}} (\mathbf{p}_{i} \mathbf{p}_{i} \transpose)$
\end{algorithmic}
\end{algorithm}

\begin{figure} [ht]
\begin{center}
\includegraphics[width= 0.6 \linewidth]{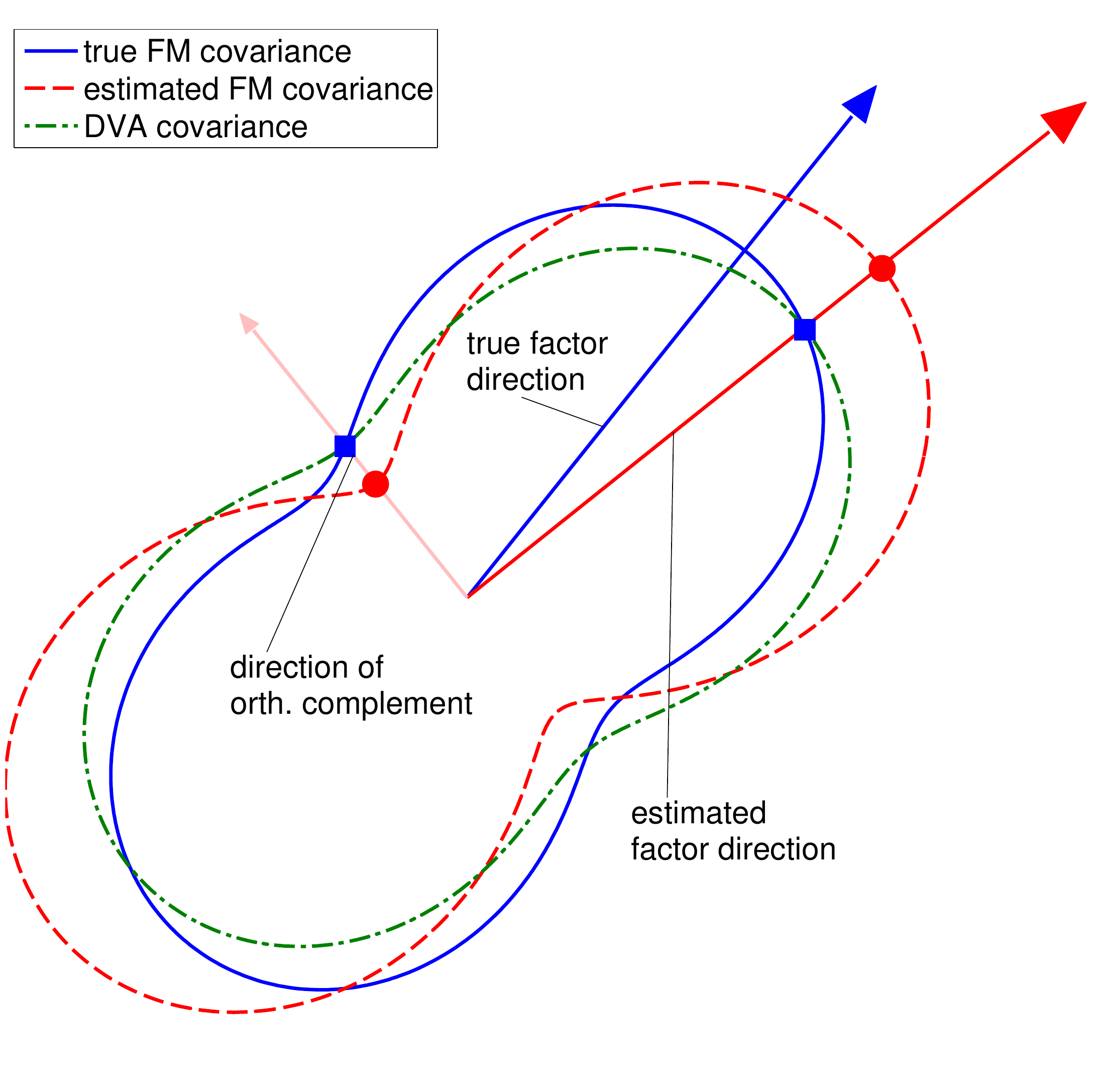}
\caption{The panel shows the directional variances for an estimated
Factor model covariance matrix (red/dashed) and the true Factor model
covariance matrix (blue/solid). The blue dots indicate the true variance
along the estimated factor direction and the direction of the
orthogonal complement. The DVA method (green/dash-dotted) aims at stretching and
compressing the estimated covariance peanut such that the variances in
these directions correspond to the true ones.} \label{fig:favarcorr}
\end{center}
\end{figure}

Note that the algorithm does not correct the parameters of the factor model
itself. Instead, only the resulting covariance matrix is adjusted. In
particular, the factor directions, i.e., the exposures, are kept
unchanged.  An illustration of an adjusted covariance matrix can be
found in Figure~\ref{fig:favarcorr}.
The figure shows in blue/solid and red/dashed the
covariances of the true and the estimated factor model,
respectively. The arrows indicate the factor
directions of the true and estimated factor model and the direction of the orthogonal
complement, respectively. Clearly, the factor direction has been misestimated and
its strength is overestimated. In the orthogonal direction the
variance is underestimated. Our proposed DVA method corrects the
systematic error of the directional variance along those directions,
without adjusting the directions themselves. This leads to the
directional variance adjusted covariance matrix (depicted in green/dash-dotted):
In the aforementioned directions, the systematic error is reduced.

One has to keep in mind that the resampling -- and with it the estimate of the systematic error of the covariance matrix -- is based on the estimated parameters $\cal F$.
Therefore, large errors in $\cal F$ adversely
affect the DVA covariance estimate. 

In order to reduce the impact of the error in $\cal{F}$, it could be advantageous to iterate the DVA procedure. From the DVA covariance matrix, which more closely reflects the true covariance matrix, we could estimate the parameters of a new factor model and restart the DVA procedure, obtaining more precise estimates of correction factors in each iteration. Though a compelling idea, there is no guarantee that iterating the DVA method will give a better solution, converge to a  sensible one or even converge at all. In this paper, we therefore concentrate on the non-iterated DVA procedure.

\subsection{Simulation Results}
Before we present results from daily return data, we will first
illustrate the effectiveness of the proposed DVA method in a
simulation study. For this, we generate toy data according to the scheme
presented in section \ref{sec:fabias}, first apply standard Factor
Analysis and then use our proposed DVA method to reduce the bias.

\begin{figure} 
  \begin{center}
    \includegraphics[width=0.48\linewidth]{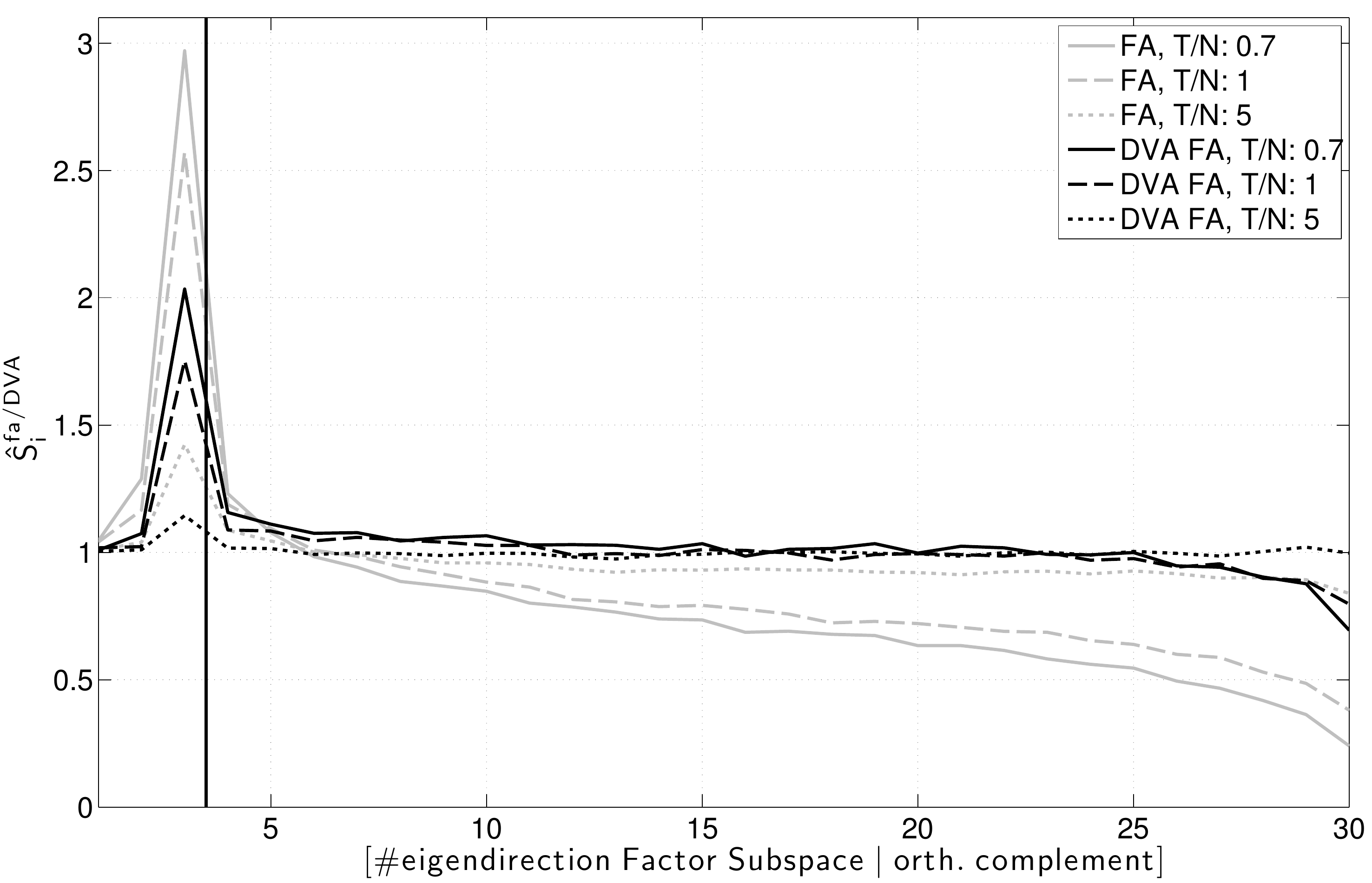}\includegraphics[width=0.48\linewidth]{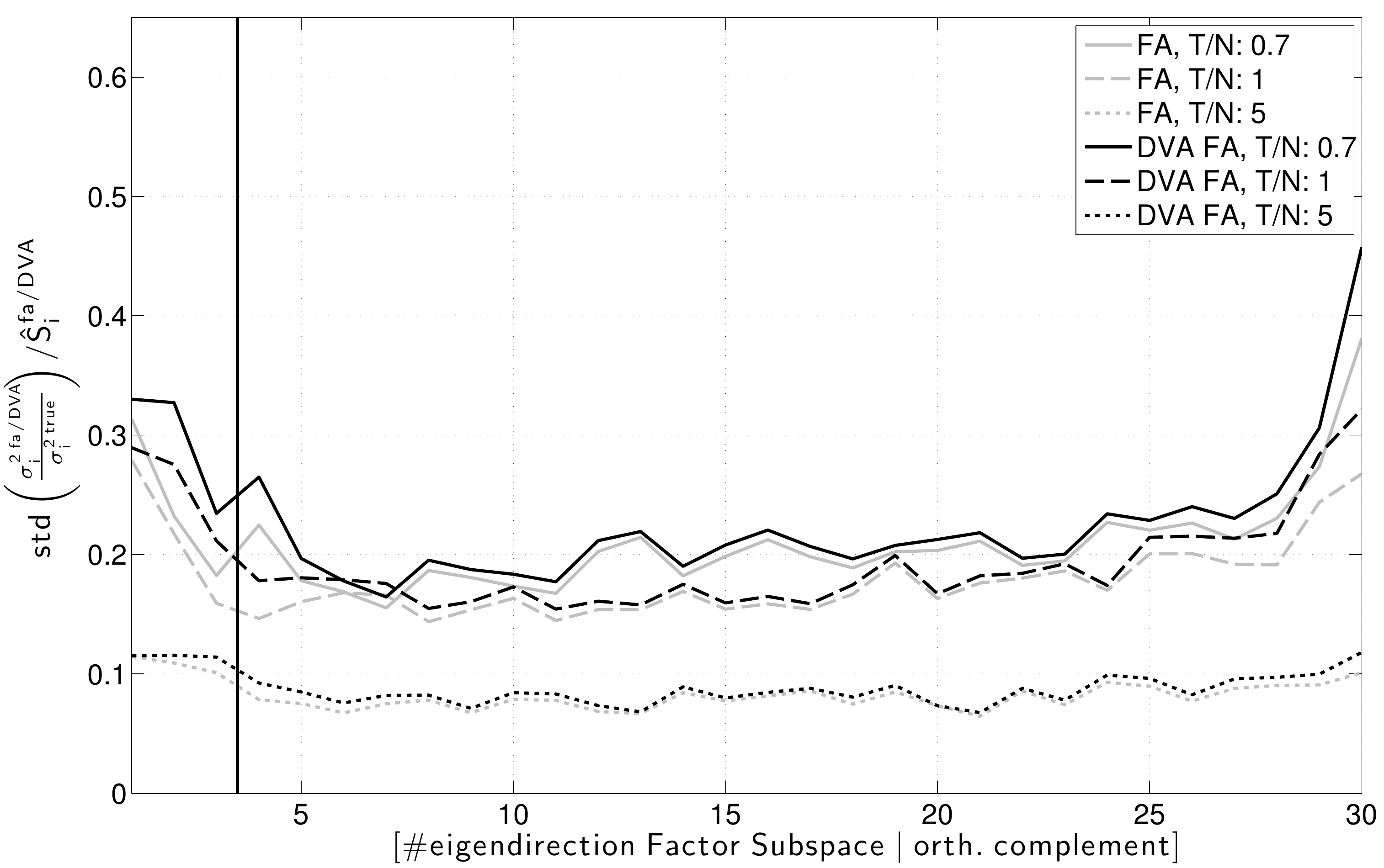}
    \caption{left: comparison of the systematic error for standard
      Factor Analysis and the DVA Factor Analysis. Right: normalized standard deviation of the error. Simulations  for different ratios of sample
      size to dimensionality ($T/N=0.7$, $1$ and $5$). $N = 30$. Correction
      factors estimated on $ K=100 $ generated data sets. Mean over
      150 simulations.} \label{fig:syscorr2}
  \end{center}
\end{figure}

The performances of the two estimation methods with respect to the
systematic error $S$ (eq.~(\ref{eq:syserr})) are contrasted in
Figure~\ref{fig:syscorr2}. To the left, it is shown that the DVA
method clearly reduces the systematic error of the Factor Analysis
model, even for relatively large ratios $T/N$. In the direction of the
third factor, which has the lowest SNR, the reduction is most
prominent. In the orthogonal complement of the factor subspace, the
adjusted spectrum resembles the true variances very
well. Nevertheless, there remains a small systematic error, which is
due to to using the \emph{estimated} parameter set  in order to infer the directional variance correction factors. The right panel of Figure \ref{fig:syscorr2} illustrates that the DVA
method does not incur a significant increase in variance of the estimate.

\begin{figure} 
  \begin{center}
    \includegraphics[width=0.8\linewidth]{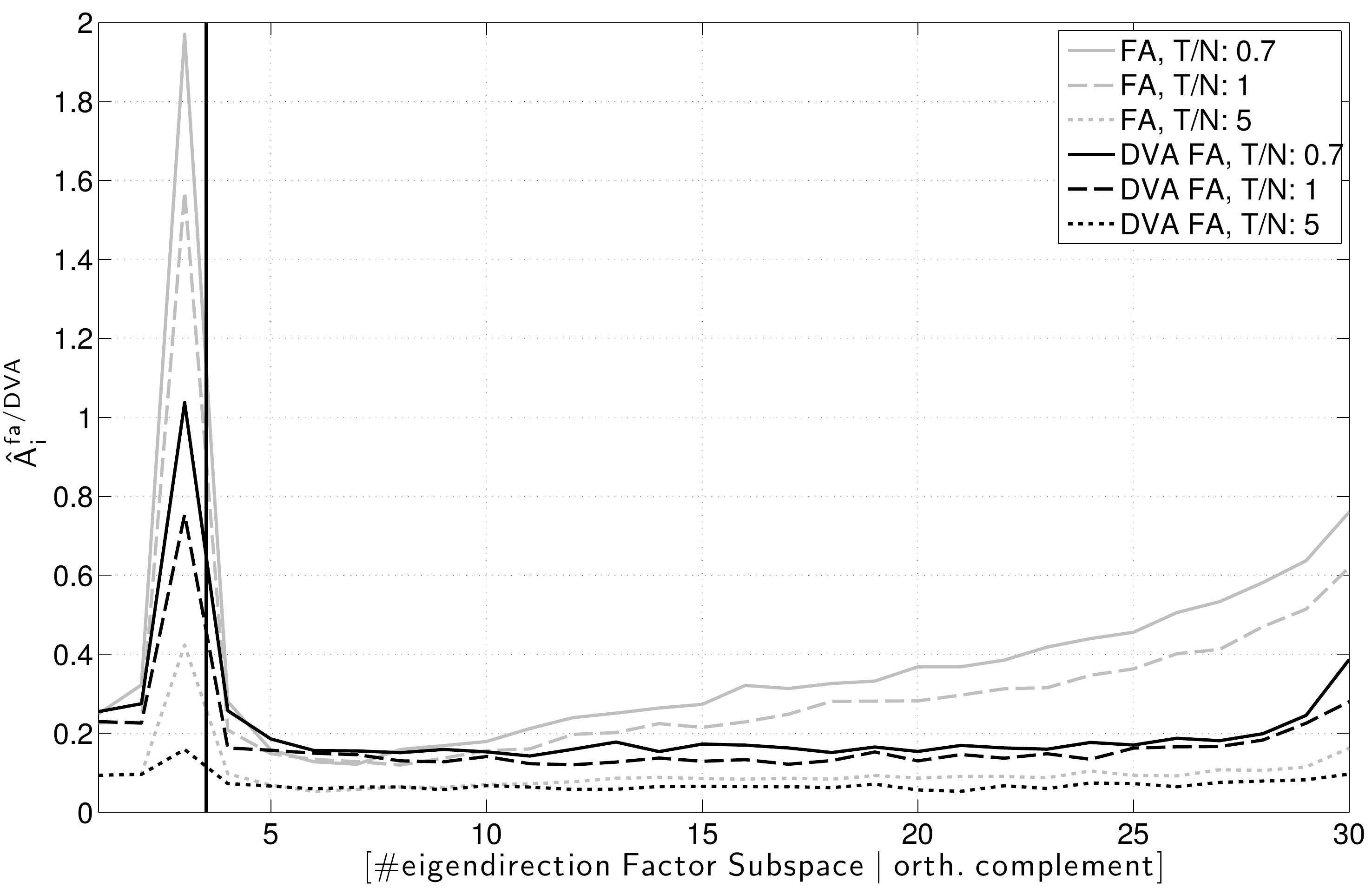}
    \caption{comparison of the mean absolute relative error for
      standard Factor Analysis and the DVA Factor Analysis for
      different ratios of sample  size to dimensionality ($T/N=0.7$, $1$ and $5$). $N = 30$. Correction factors estimated on $ K=100 $ generated
      data sets. Mean over 150 simulations.} \label{fig:syscorrerr2}
  \end{center}
\end{figure}

By reducing the systematic error without an increase in variance, the
DVA method reduces the average estimation error. To account for
different magnitudes of true directional variances,
Figure~\ref{fig:syscorrerr2} displays the error of the
estimator in terms of the mean absolute relative error
\begin{align} 
A^{fa / DVA}_i = E \left[ \frac{ | \sigma^{2 \, fa / DVA}_i -
    \sigma^{2 \, true}_i |}{\sigma^{2 \, true}_i} \right].
\end{align} 
Note that this error is more than halved for the direction of the low
SNR-factor and considerably decreased in the orthogonal
complement. Here, DVA has the strongest effects on the directions
corresponding to the largest and smallest non-zero eigenvalues of
$\mathbf{C}^{fa}_{oc}$. For the direction of the smallest eigenvalue,
the error is again approximately halved.

\begin{figure} 
  \begin{center}
    \includegraphics[width=0.8\linewidth]{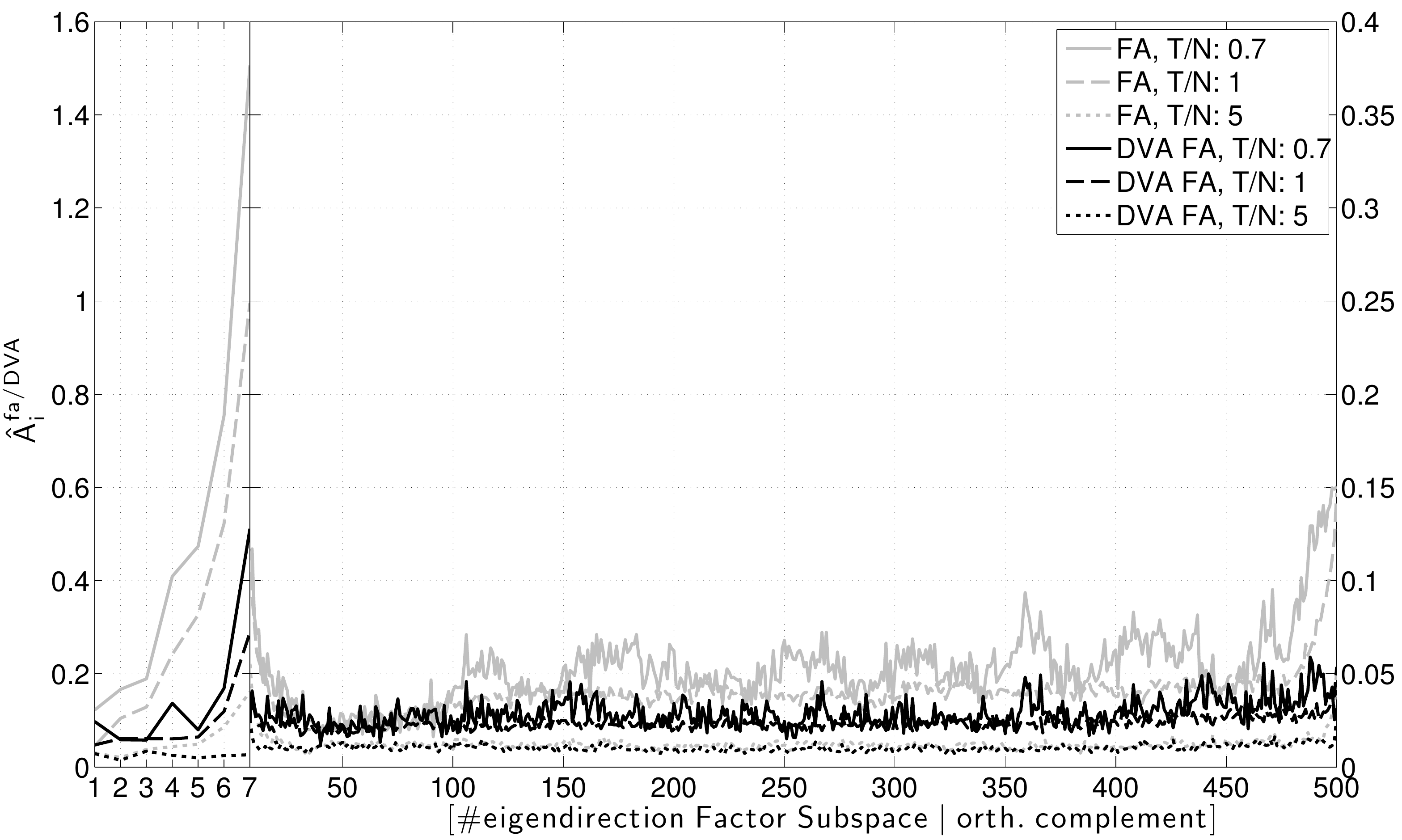}
    \caption{comparison of the mean absolute relative error for
      standard Factor Analysis and the DVA Factor Analysis for
      different ratios of sample  size to dimensionality ($T/N=0.7$, $1$ and $5$). Note that the y-axis has different scaling for the factor subspace and the orthogonal complement. $N = 500$. Correction factors estimated on $ K=100 $ generated
      data sets. Mean over 150 simulations.} \label{fig:syscorrerr2HD}
  \end{center}
\end{figure}

While the ratio $T/N$ determines most properties of the sample covariance, this is not true for regularized estimators and factor models. For larger values of $T$, at a constant ratio $T/N$, the idiosyncratic variances of Factor Analysis are estimated more precisely, while the estimation of the factors remains difficult. This is shown in Fig.\,\ref{fig:syscorrerr2HD}, where the dimensionality has been set to 500 and the generative model has seven factors of strength 10, 5, 4, 3, 2.5, 2, 1.5, and 1. One can see that while there is little room for improvement in the orthogonal complement,  in the Factor Subspace the performance gain by DVA FA remains on the same level.

\section{Empirical Results}
\label{sec:empirics}

\subsection{Portfolio Simulation}
In order to evaluate the proposed methods, we applied the DVA Factor
Analysis to financial daily return time series. In the experiments, we
estimate covariance matrices of stock returns and use the covariance
estimates for portfolio optimization. The realized risks of the
portfolios are compared for the different covariance estimates. In
particular, we will compare the DVA Factor Analysis to the sample
covariance matrix, Resampling Efficiency\footnote{Although Resampling Efficiency does not yield a covariance estimate, we include it in the comparison.}, the Fama-French Three-Factor model (see,
e.\,g.,  \cite{Fan08}, \cite{FamFre92}), Shrinkage to a one-factor model (\cite{LedWol03}) and standard Factor Analysis (see
section \ref{sec:fa}). For DVA and standard Factor Analysis we use seven factors. Though on the higher dimensional US and EU data sets we could extract more factors and fewer factors would be favorable on the smaller HK data set, we opted for the same intermediate model complexity on all data sets to keep the setting simpler.

\subsection{The Data Sets}
The data set consists of daily returns of about 1300 US stocks
(3.1.2001--2.11.2009), about 600 European stocks (3.1.2001--20.4.2009)
and a set of 200 stocks from the Hong Kong stock exchange
(3.1.2001--26.9.2008). Removing stocks which do not have data for the
whole time horizon covered by the data set, the Hong Kong data set
reduces to 100 assets.

\subsection{Design of Portfolio Simulations}
There are different applications of covariance matrices in portfolio
optimization. Covariance matrices are needed for index tracking,
hedging and the search for minimum variance portfolios. In the
following, we will focus on minimum variance portfolios. The minimum
variance portfolio is given by
\begin{align}
\mathbf{w}^* = 
\mathrm{arg}\!\min_\mathbf{\hspace{-1.5em}w} \;
\mathbf{w} \transpose \hat{\mathbf{C}} \ \mathbf{w},
\label{eq:minvar}
\end{align}
where $\mathbf{w}$ is the vector of portfolio weights and
$\hat{\mathbf{C}}$ is the covariance matrix estimate.

Depending on the particular application, additional constraints are
incorporated into the optimization. Commonly applied constraints
include:
\begin{itemize}
 \item $\sum_i w_i = 1$: the sum of all portfolio weights is
 restricted to one.  
 \item $ \mathbf{w} \transpose \hat{\mathbf{r}} = \mathrm{r}^*$:
   the estimated portfolio return is restricted to $\mathrm{r}^*$,
   $\hat{\mathbf{r}}$ is the vector of expected/predicted asset
   returns.  \item $w_i \geq 0$: only positive portfolio weights, no
   short-selling.
\end{itemize}

Note that the application of constraints tremendously prunes the set
of feasible portfolios and hence diminishes the influence of the
covariance estimate (for details, see \cite{JagMa03}). Consequently,
the observed differences between the performances of portfolios
obtained from different covariance estimation methods get smaller. Thus, in
order to unveil the leverage of the various covariance estimation methods, we
opted for not constraining the magnitude of the weights or enforcing
their positivity. We only applied the constraint that scales the sum
of the portfolio weights to one\footnote{This optimization is
  independent of the return estimates and is equivalent to optimizing
  portfolio returns under the assumption of equal expected returns for
  all assets.}. In the case of small sample sizes, this approach will
tend to overfit the directions of smallest variance and is hence expected
to favour the restricted covariance estimators. Therefore, in section \ref{sec:addreg}
we also investigate the performances of portfolios obtained
from a regularized optimization problem of
eq.~(\ref{eq:minvar}), where the additional regularization enforces
diversified portfolios.

In order to evaluate the performance of the different covariance
estimator we use the realized (out-of-sample) variance of the
estimated portfolios:
\begin{equation}
 \sigma^2_{real} = \frac{1}{T} \sum_{t=1}^T \left[ \mathbf{w}_{t-1} \transpose
  (\mathbf{r}_t - \hat{\mathbf{r}}_{t-1}) \right]^2,
 \label{eq:realVar}
\end{equation}
and, of more financial interest, the realized mean absolute deviation
\begin{equation}
 \text{MAD}_{real} = \frac{1}{T} \sum_{t=1}^T \left| \mathbf{w}_{t-1}
 \transpose (\mathbf{r}_t - \hat{\mathbf{r}}_{t-1}) \right|.
 \label{eq:realMAD}
\end{equation}
Note, that \eqref{eq:realVar} and \eqref{eq:realMAD} are rolling out-of-sample estimates, as $\mathbf{w}_{t-1}$ and $\hat{\mathbf{r}}_{t-1}$ are the
portfolio weights and expected returns estimated on the information
available until time $t-1$. More precisely, for the estimation of the
covariance matrix $\hat{\mathbf{C}}_{t-1}$ and the averaged return
$\hat{\mathbf{r}}_{t-1}$ we used a strictly causal window of 150
trading days.

In order to reduce the variance of the performance evaluation and to
thoroughly explore the estimated covariance structure, $J = 1000$
subsets, each confined to 40 (HK) or 100 (US and EU) assets, are chosen and the optimal
(confined) portfolio $\mathbf{w}^j_t$ is constructed from the given
covariance matrix estimate $\hat{C}_t^j$. The realized variance and realized absolute deviation are then
determined based on the average performance across the different confined
portfolios, i.e.,
\begin{align}
  \sigma^2_{real}  =  \frac{1}{T} \sum_{t=1}^T \left\{ \frac{1}{J} \sum_{j=1}^{J} \left[ ({\mathbf{w}_{t-1}^j})
 \transpose (\mathbf{r}_t - \hat{\mathbf{r}}_{t-1}) \right]^2 \right\}, \notag \\
 \text{MAD}_{real} =  \frac{1}{T} \sum_{t=1}^T \left\{ \frac{1}{J} \sum_{j=1}^{J} \left| ({\mathbf{w}_{t-1}^j})
 \transpose (\mathbf{r}_t - \hat{\mathbf{r}}_{t-1}) \right| \right\}. \notag
\end{align}

\subsection{Results and Discussion of Portfolio Simulations}
In this section we will provide portfolio simulation results for
different covariance estimation approaches, namely the sample
covariance matrix, Resampling Efficiency, the Fama-French three-factor model, Shrinkage to a one-factor Model, a Factor
Analysis model with seven factors, and a directional variance adjusted
Factor Analysis (DVA FA, section~\ref{sec:fabias}).  The results for
the different markets are summarized in Table~\ref{tab:unregResults}.

\begin{table}[b!]
  \begin{center}
    \begin{tabular}{ c | c | c | c }
      & US & EU & HK \\
      \hline
Sample Cov. & 8.56$^{\dagger}$ (156.1$^{\dagger}$) & 5.93$^{\dagger}$ (78.9$^{\dagger}$) & 6.57$^{\dagger}$ (81.2$^{\dagger}$) \\
Resampling Eff. & 8.83$^{\dagger}$ (165.7$^{\dagger}$) & 6.11$^{\dagger}$ (83.5$^{\dagger}$) & 6.64$^{\dagger}$ (82.7$^{\dagger}$) \\
Fama-French & 5.65$^{\dagger}$ (73.5$^{\dagger}$) & 3.97$^{\dagger}$ (38.6$^{\dagger}$) & 6.20$^{\dagger}$ (73.4$^{\dagger}$) \\
LW Shrinkage & 5.56$^{\dagger}$ (69.6$^{\dagger}$) & 4.00$^{\dagger}$ (39.1$^{\dagger}$) & 6.17$^{\dagger}$ (72.9$^{\dagger}$) \\
Factor Analysis & 5.47$^{\dagger}$ (67.8$^{\dagger}$) & 3.88$^{\dagger}$ (36.5$^{\dagger}$) & 6.17$^{\dagger}$ (73.0$^{\dagger}$) \\
DVA FA & \textbf{5.40}$^{}$ (\textbf{66.7}$^{}$) & \textbf{3.84}$^{}$ (\textbf{36.0}$^{}$) & \textbf{6.12}$^{}$ (\textbf{71.7}$^{}$) \\
\end{tabular}
    \caption{Mean absolute deviations$\cdot 10^3$ (mean squared
      deviations$\cdot 10^6$) of the resulting portfolios for the
      different covariance estimators and the different markets. $^\dagger:=$ DVA mean significantly better/worse than this
      model at the 5\% level, tested by a randomization
      test.}  \label{tab:unregResults}
  \end{center} 
\end{table}

As expected, the sample covariance matrix is not the most suitable
tool for portfolio optimization. Across all data sets, the portfolios
derived from the different factor based models and Shrinkage clearly outperform the
sample covariance matrix based portfolios in terms of realized risk. A
direct comparison of these models reveals that
the DVA method always significantly outperforms Fama-French, standard Factor
Analysis and Shrinkage with respect to realized variance and realized absolute
deviation. On our data sets, Resampling Efficiency does not give an advantage over the sample covariance matrix.


\subsection{Results and Discussion of Portfolio Simulations -- Additional Regularization}
\label{sec:addreg}
Without knowledge of the covariance structure of the assets, the best
portfolio allocation would have weights inverse to the variance of the
assets and hence be highly diversified.  Minimization of
eq.~(\ref{eq:minvar}), on the other hand, gives the optimal portfolio
only for the true covariance matrix. Therefore, for a given 
covariance matrix estimate, it should in principle be possible to additionally
reduce the realized risk of a portfolio by increasing its
diversification, e.g., by regularization of eq.~(\ref{eq:minvar}).

Consequently, the aim of the following analysis is twofold. First of all
and from a theoretical perspective, we want to investigate if the
superior performance of the DVA method can be simply explained away by
a higher degree of diversification or if the true covariance structure
is indeed better captured. Secondly, with respect to practical
considerations, we are interested in the best achievable performance.

In order to analyze these aspects, for each of the  covariance
matrix estimates $\hat{\mathbf{C}}$ we enforce additional portfolio
diversification by including a ridge penalty in the objective function
eq.~\eqref{eq:minvar}, i.e.,
\begin{align}
\mathbf{w}^*(\lambda) = 
\mathrm{arg}\!\min_\mathbf{\hspace{-1.5em}w} \;
\mathbf{w} \transpose \hat{\mathbf{C}} \ \mathbf{w} + \lambda \mathbf{w} \transpose \mathbf{\Lambda}\  \mathbf{w} .
\label{eq:regMinVar}
\end{align}
In particular, we set the metric $\mathbf{\Lambda}$ to a diagonal
matrix which has the single asset variances on its diagonal. This metric
implies that each asset gets penalized by its variance and in the
limit $\lambda \rightarrow \infty$ we obtain the portfolio of assets
weighted by the inverse of their variances.

\begin{table}[b!]
  \begin{center}
    \begin{tabular}{ c | c | c | c }
      & US & EU & HK \\
      \hline
Sample Cov. & 5.45$^{\dagger}$ (67.3$^{\dagger}$) & 3.91$^{\dagger}$ (37.0$^{\dagger}$) & 6.14$^{\dagger}$ (72.8$^{}$) \\
Resampling Eff. & 5.48$^{\dagger}$ (67.7$^{}$) & 3.93$^{\dagger}$ (37.2$^{}$) & 6.16$^{\dagger}$ (73.4$^{}$) \\
Fama-French & 5.55$^{\dagger}$ (70.0$^{\dagger}$) & 3.93$^{\dagger}$ (37.7$^{}$) & 6.10$^{}$ (71.6$^{}$) \\
LW Shrinkage & 5.39$^{\dagger}$ (65.8$^{}$) & 3.86$^{\dagger}$ (36.3$^{}$) & 6.10$^{}$ (71.8$^{}$) \\
Factor Analysis & 5.38$^{\dagger}$ (66.0$^{}$) & 3.82$^{\dagger}$ (35.6$^{}$) & 6.09$^{}$ (71.7$^{}$) \\
DVA FA & \textbf{5.35}$^{}$ (\textbf{65.6}$^{}$) & \textbf{3.81}$^{}$ (\textbf{35.5}$^{}$) & \textbf{6.09}$^{}$ (\textbf{71.3}$^{}$) \\
\end{tabular}
    \caption{Mean absolute deviations$\cdot 10^3$ (mean squared
      deviations$\cdot 10^6$) of the resulting portfolios for the
      different regularized covariance estimators for optimal regularization strength and the different markets.  $^\dagger:=$ DVA mean significantly better/worse than this
      model at the 5\% level, tested by a randomization
      test. }  \label{tab:regResults}
  \end{center} 
\end{table}

\begin{figure}
  \begin{center}
    \includegraphics[width=.49\textwidth]{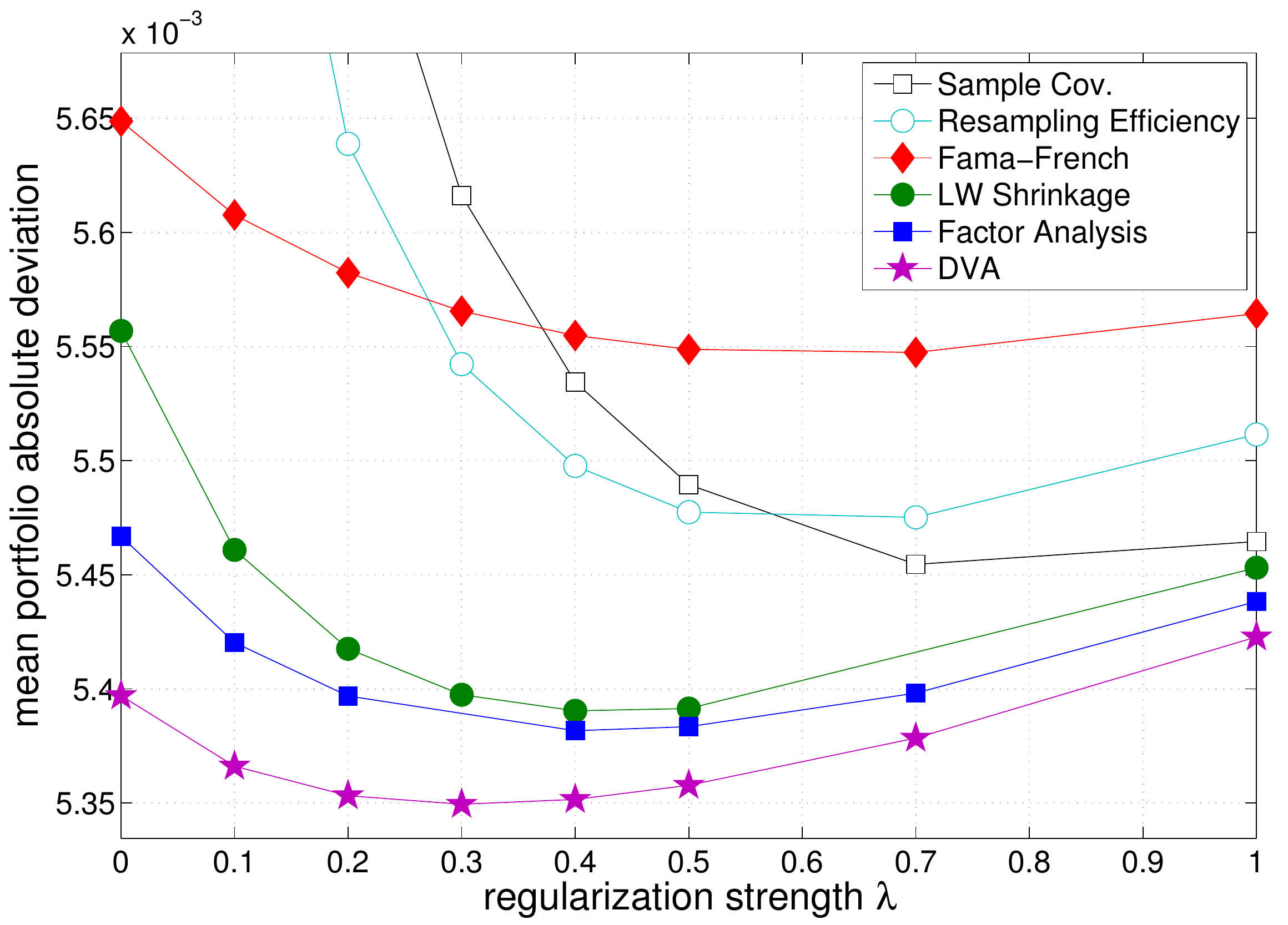}
    \includegraphics[width=.49\textwidth]{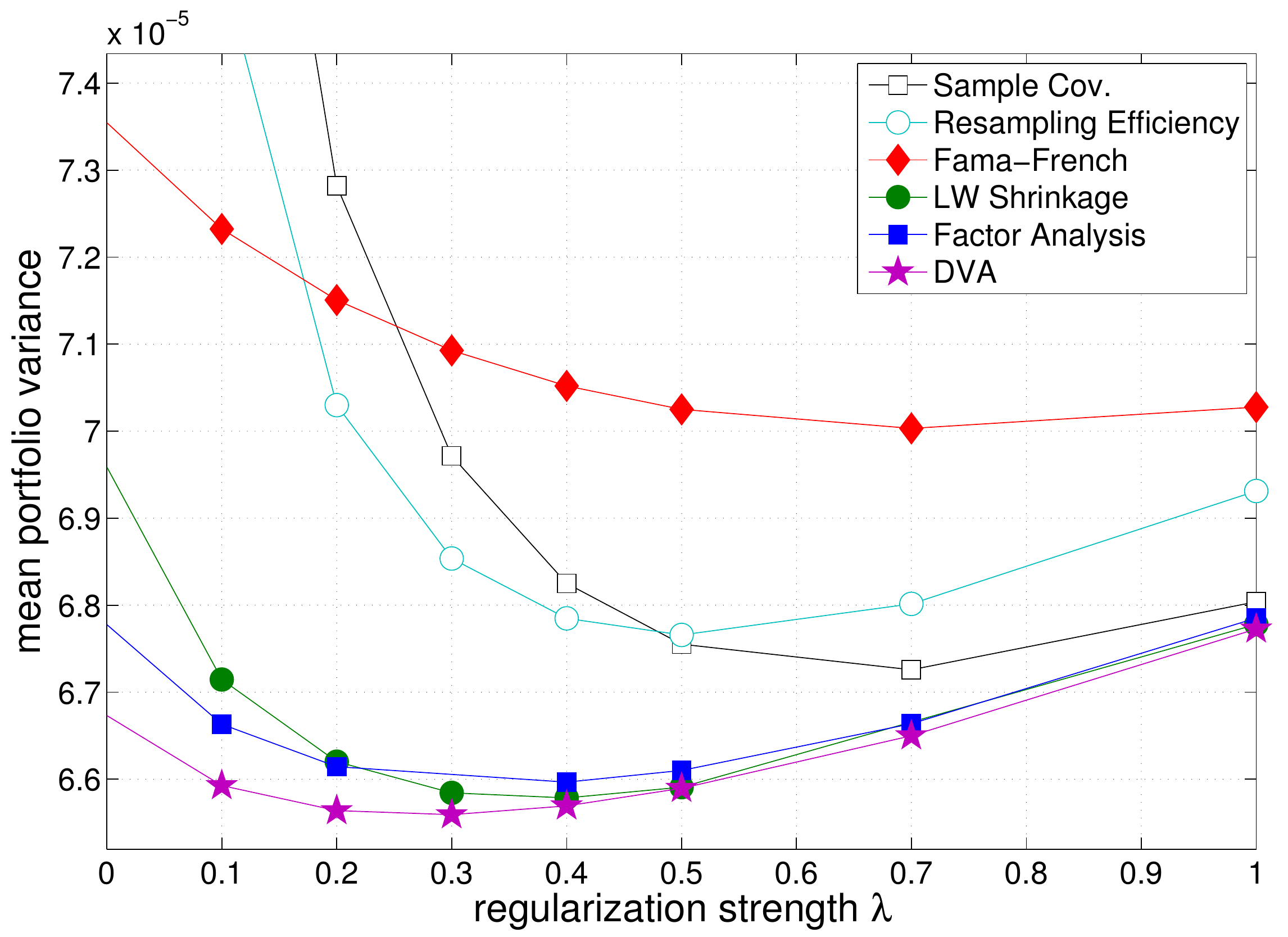}
    \caption{Realized portfolio risk. Left: mean absolute
      deviation. Right: variance. US market.}
 \label{fig:USt100_reg}
  \end{center}
\end{figure}

\begin{figure} 
  \begin{center}
    \includegraphics[width=.49\textwidth]{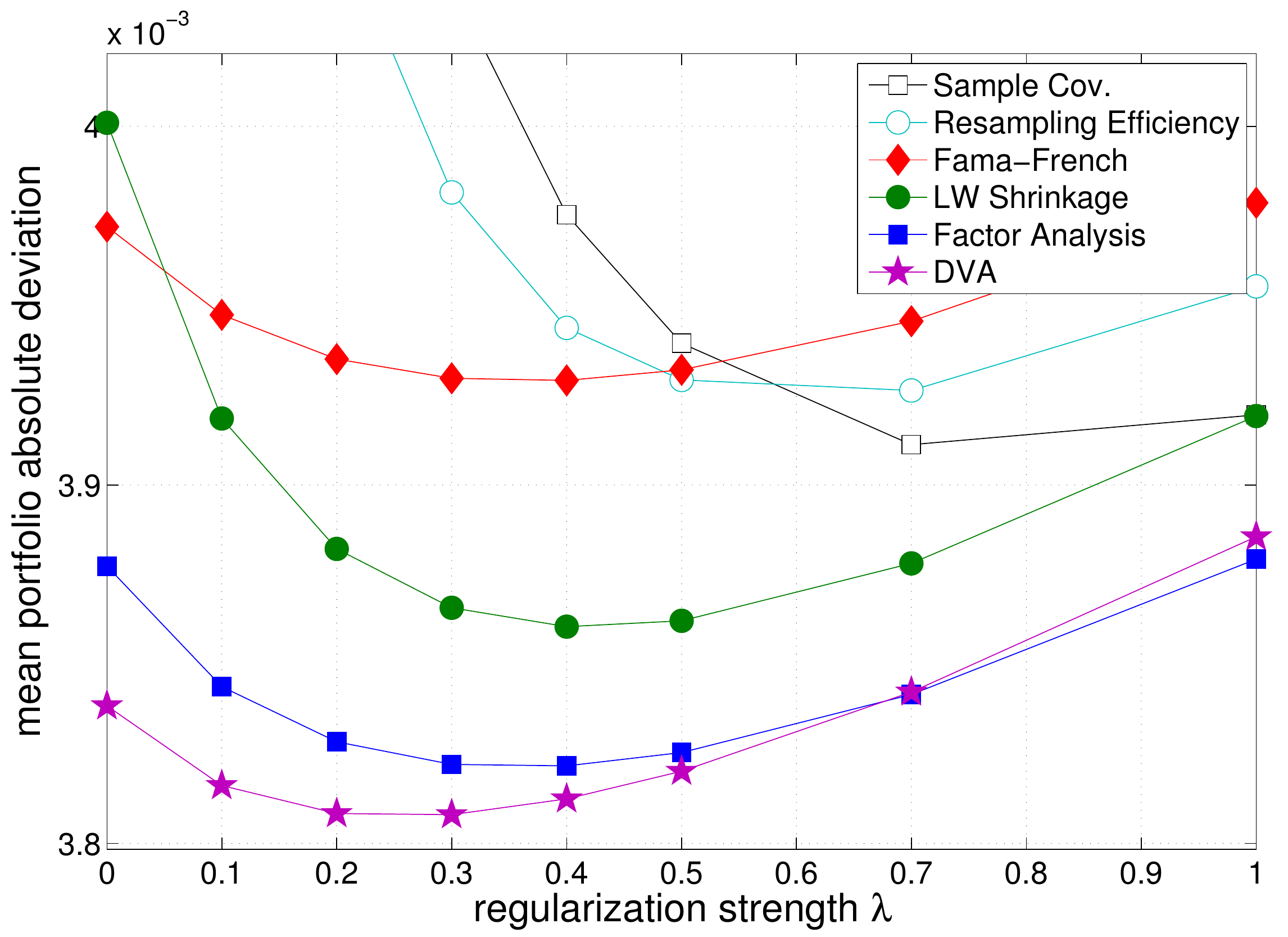}
    \includegraphics[width=.49\textwidth]{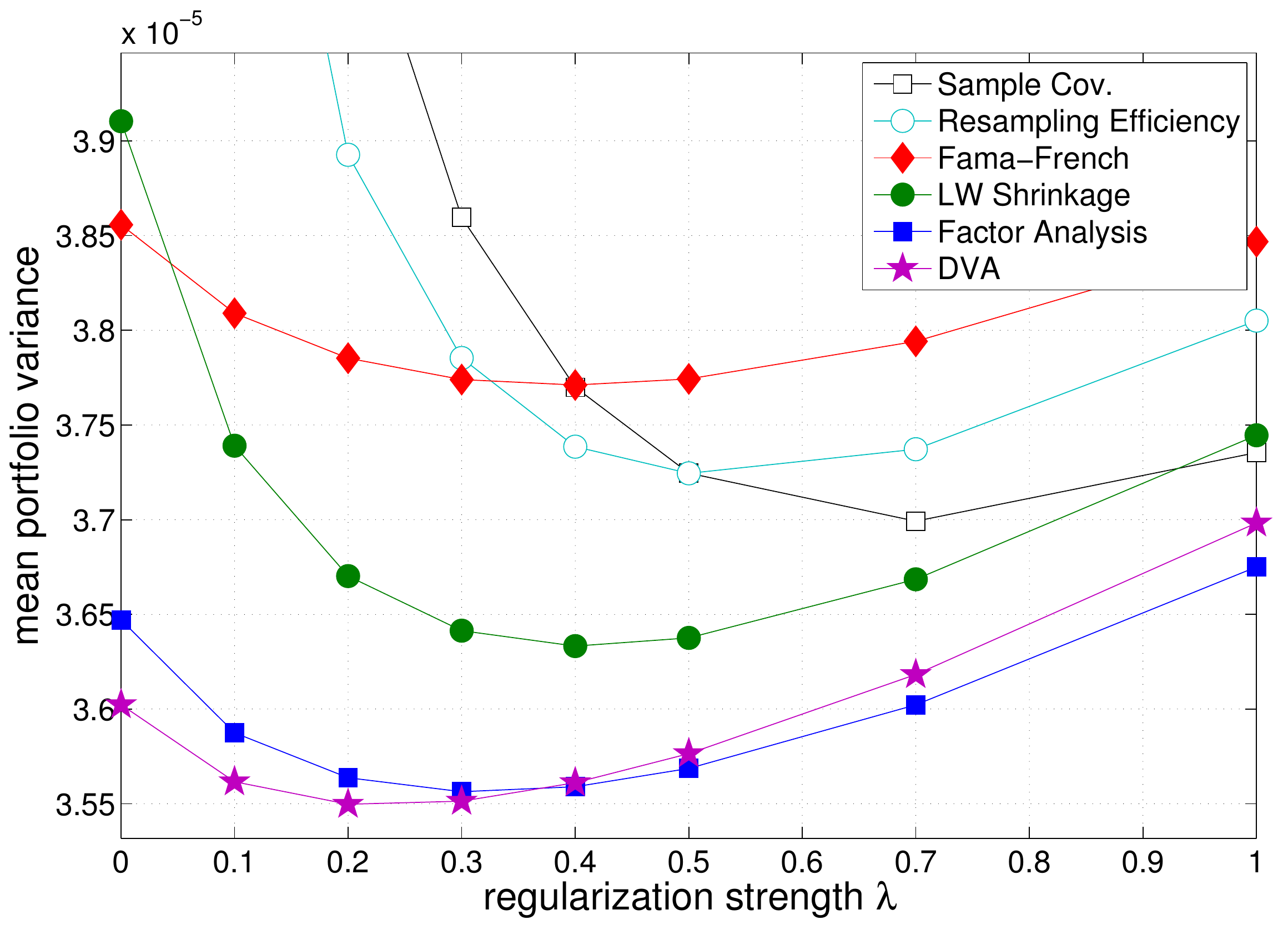}
    \caption{Realized portfolio risk. Left: mean absolute
      deviation. Right: variance. EU market.}
    \label{fig:USb_reg}
  \end{center}
\end{figure}

\begin{figure} 
  \begin{center}
    \includegraphics[width=.49\textwidth]{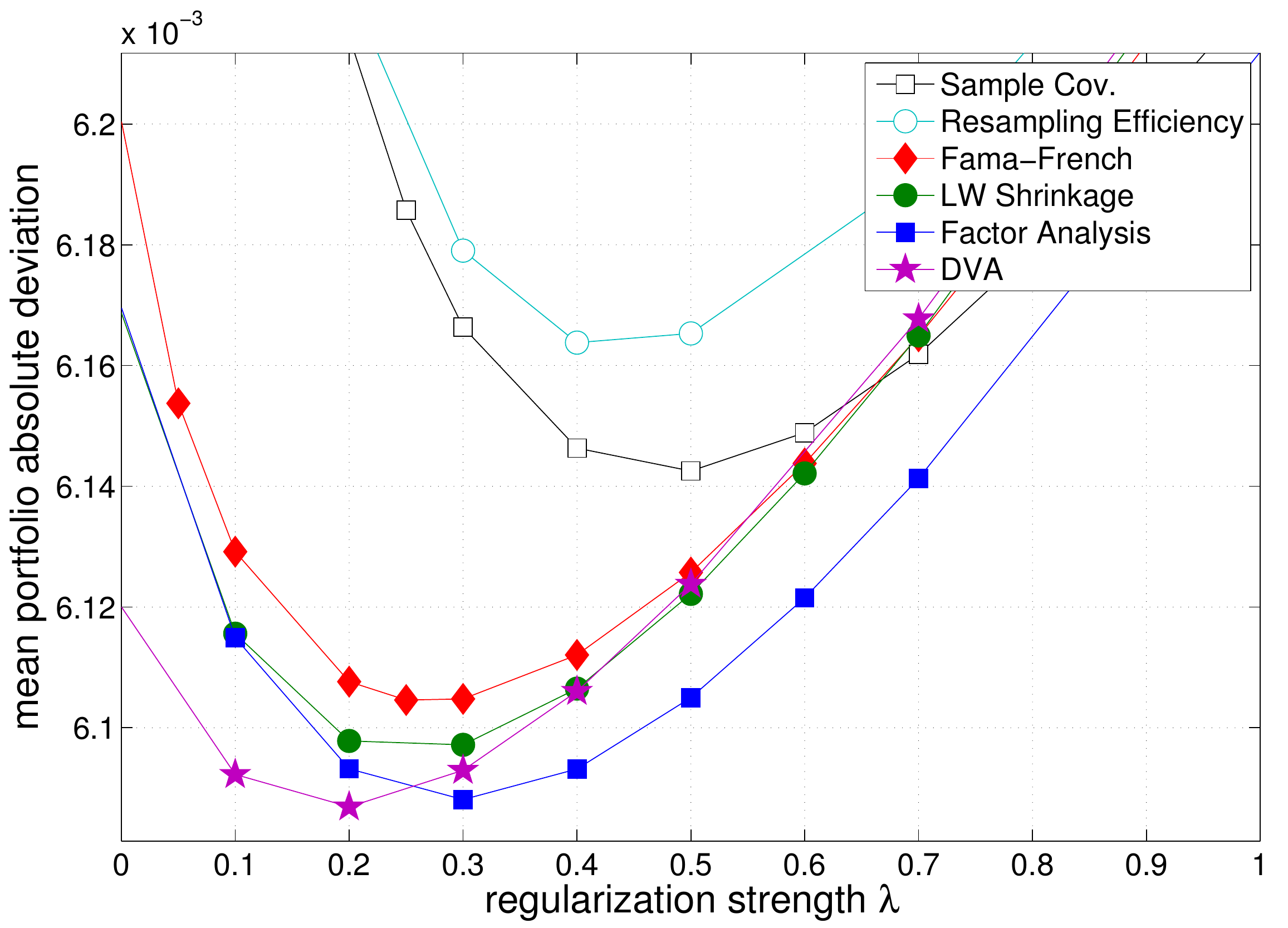}
    \includegraphics[width=.49\textwidth]{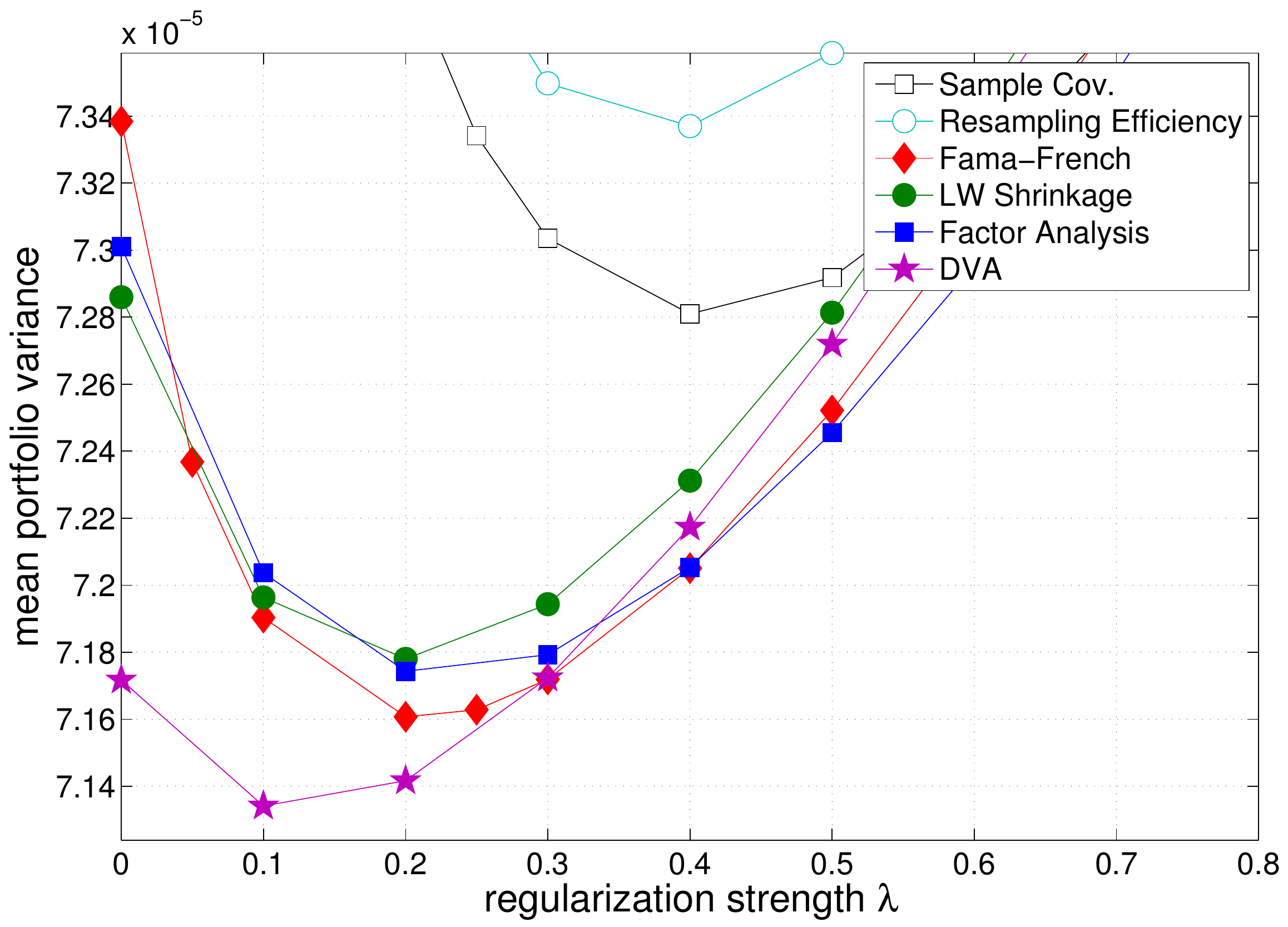}
    \caption{Realized portfolio risk. Left: mean absolute
      deviation. Right: variance. HK market.}
    \label{fig:HK_reg}
  \end{center}
\end{figure}

Figure~\ref{fig:USt100_reg} -- \ref{fig:HK_reg} depict the realized
(out-of-sample) variance and MAD (see eq.\,\eqref{eq:realVar} and
eq\,\eqref{eq:realMAD}) of the resulting portfolios as a function of the
regularization parameter $\lambda$ for the three different market samples.

In unison, the different models benefit from additional
regularization, as can be seen from a reduction of the realized risk
of the resulting portfolios (cmp.\ Tables \ref{tab:unregResults} and \ref{tab:regResults}). Although, this effect is most pronounced
for the sample covariance matrix, it merely reaches the performance of
the (unregularized) Factor Analysis models. Note that the regularized
optimization based on the sample covariance matrix is equivalent to
unregularized optimization using a shrinkage covariance estimator,
that employs ${\mathbf{C}}^{target} = \mathbf{\Lambda}$ as the
shrinkage target (cf.~eq.~(\ref{eq:shrinkage})). Again, Resampling Efficiency does not prove to be superior to the sample covariance matrix.

Shrinkage to the one-factor model profits as well from additional Shrinkage to  $\mathbf{\Lambda}$. This indicates that the optimization of  the expected mean squared error of automatic Shrinkage gives a too small Shrinkage parameter for the optimization of portfolios.

Surprisingly, the Fama-French Three-Factor model does not benefit as
much as Shrinkage from the regularization, although the unregularized performance is similar. 
This implies that the performance gain of the unregularized Fama-French model over the sample covariance matrix is mainly due to a strong imposed
prior towards highly diversified portfolios. 
Compared to the statistical FMs FA, and DVA FA, the performance difference remains on the same level as without additional regularization. 
This means that the covariance structure is better captured by the statistical FMs than by the
Fama-French model.
These effects are strongest for the US and EU markets.

The risk of the portfolios obtained from the Factor Analysis model as
well as from its DVA version also improve considerably. 
At the
optimal degree of regularization, the DVA FA model significantly
outperforms the optimally regularized sample covariance matrix based
model for all markets. Regarding eq.\,\eqref{eq:regMinVar} as being a shrinkage towards
$\mathbf{\Lambda}$, this statement is equivalent to: shrinkage of
the DVA Factor Analysis covariance matrix towards $\mathbf{\Lambda}$
yields better portfolios with respect to the achieved portfolio risks
than shrinkage of the sample covariance matrix towards
$\mathbf{\Lambda}$. 
The comparison of FA DVA with  Fama-French shows a significantly better performance for all markets as well. The performance gain over Shrinkage is, however, only significant for US and EU markets.

At the optimal degree of regularization the difference in performance
between the standard Factor Analysis and the DVA Factor Analysis is
reduced. In general, this was to be expected as regularization can
equivalently be achieved either by adding a penalty term to the
objective function or by additionally constraining the feasible set.
In this respect, it was shown in \cite{JagMa03} that the actual
influence of the  covariance matrix estimate on the minimum
variance portfolio diminishes when additionally constraining the set
of feasible portfolios.  Thus, as a matter of fact, regularization
partly compensates for the influence of the systematic error of the
Factor Analysis covariance matrix estimate. 

Nevertheless, in the US
and EU market, the difference in mean MAD remains
significant at the 5\% level. In Hong Kong the peformance gain of DVA over standard Factor Analysis  is, for optimal regularization, not significant.

Comparing the different markets, it turns out that the Hong Kong
market shows a slightly different behavior than the American and
European. At the Hong Kong market, all methods likewise benefit from
additional diversification. One possible explanation is that the HK
data set contains quite a few outliers and missing data as opposed to
the US and EU data. Thus covariance estimates as well as least square
estimates of factor exposures are hampered in general. Hence and in
contrast to the other markets, the Fama-French model also clearly
profits from the additional regularization, although its overall
performance remains inferior to DVA Factor Analysis.


\section{Summary}
\label{sec:discussion}

The fundamental issue in portfolio allocation is the accurate and
precise estimation of the  covariance matrix of asset returns
from historical data. Among many challenges, the data is typically
high dimensional, noisy, contaminated with outliers and
nonstationarity interferes with the use of long estimation
windows. Thus, reliable statistical parameter estimation is often
impeded. Our work has contributed to alleviate this problem in
theoretical and practical aspects: (1) we demonstrated that the data
driven statistical Factor Analysis model has a systematic estimation
error, which can be alleviated by the proposed algorithmic Directional
Variance Adjustment (DVA) framework, (2) a DVA correction of Factor
Analysis yields substantial improvements for minimum variance
portfolios, and finally (3) extensive simulations of portfolios of EU,
US and Hong Kong markets underpinned the usefulness of the DVA
approach in terms of significant gains in realized variance
and realized mean absolute deviation.

For each covariance estimator, we additionally studied the effect of
regularizing the minimum variance portfolios towards a higher degree
of diversification. As expected, diversification improved portfolio
performance across the different estimators. Our empirical study showed
that while regularization slightly decreases the overall advantage
gained by DVA, the remaining difference in the minimum stayed
significant for the US and EU  data sets, here the DVA Factor Analysis method is superior to standard Factor
Analysis. 

A second interesting finding of the regularization experiments was
that the advantage of the Fama-French model over  the sample
covariance matrix estimator appears rather due to an imposed strong
diversification prior than to an improved estimation of the underlying
covariance structure. Here, clearly the combination of regularization and 
statistical FMs like standard FA and in particular DVA FA
led to better model performance. 

Note, however, that down-weighting/regularizing away the estimated
correlations may not always be a valid option. In an application where
the covariance structure is of higher importance -- e.g. because an
index needs to be tracked with a reduced number of assets -- increased
diversification would clearly be no option.

Therefore, both scenarios, the one with and the one without
regularization, yield interesting insight and a clear gain when using
DVA FA.

Whilst we have studied and modeled daily returns, the DVA method is of
course equally capable of being employed to derive covariances for
intraday returns. Intraday covariance matrices are particularly
relevant when dealing with portfolios with significant (intraday)
churn.  Examples of such portfolios include internalization portfolios
at most major brokerages, and those used for market making. Using DVA FA, 
a covariance matrix may be tuned for the typical period a position
remains in a portfolio, allowing, potentially, better risk management
and asset allocation.

We do not consider serial correlation, as it is common for covariance estimation methods like Shrinkage (see \cite{LedWol03,LedWol04}) and statistical Factor Models (see, e.g., \cite{Con10}). Nevertheless, it would be interesting to do further research on an autoregressive Factor Analysis model.

\section{Acknowledgements}
We are grateful to Gilles Blanchard for his valuable comments. We thank two anonymous reviewers who supplied helpful suggestions which led to substantial improvements of the manuscript.


\bibliographystyle{elsarticle-harv}
\bibliography{journal_macros_short,ida,machineLearning,finance,dbML}{}







\end{document}